\documentclass{aa}  

\usepackage{graphicx}
\usepackage{amsmath}	
\usepackage{subcaption}
\usepackage{txfonts}
\usepackage[dvipsnames]{xcolor}
\usepackage[colorlinks=true,citecolor=blue]{hyperref}
\begin{document}

   \title{FLAMINGO: Galaxy formation and feedback effects on the gas density and velocity fields}

   \author{Lurdes Ondaro-Mallea
          \inst{1,2}
          ,
          Raul E. Angulo
          \inst{1,3}
          ,
          Giovanni Aricò
          \inst{4}
          , 
          Joop Schaye
          \inst{5}
          ,
          Ian G. McCarthy
          \inst{6}
          ,
          Matthieu Schaller
          \inst{7,5}
          }

   \institute{
         Donostia International Physics Center, Manuel Lardizabal Ibilbidea, 4, 20018 Donostia, Gipuzkoa, Spain\\
         \email{lurdes.ondaro@dipc.org}    
         \and 
         Department of Theoretical Physics, University of the Basque Country UPV/EHU, Bilbao, E-48080, Spain
        \and
         IKERBASQUE, Basque Foundation for Science, 48013, Bilbao, Spain
         \and
         Institut für Astrophysik (DAP), Universität Zürich, Winterthurerstrasse 190, 8057 Zürich, Switzerland
         \and
         Leiden Observatory, Leiden University, PO Box 9513, 2300 RA Leiden, the Netherlands
         \and 
         Astrophysics Research Institute, Liverpool John Moores University, Liverpool L3 5RF, UK
         \and 
         Lorentz Institute for Theoretical Physics, Leiden University, PO box 9506, 2300 RA Leiden, the Netherlands
            }
    \titlerunning{Density and velocity fields of gas}
   \authorrunning{L. Ondaro-Mallea et al.}


 
  \abstract{Most of the visible matter in the Universe is in a gaseous state, subject to hydrodynamic forces and galaxy formation processes that are much more complex to model than gravity. These baryonic effects can potentially bias the analyses of several cosmological probes, such as weak gravitational lensing. In this work, we study the gas density and velocity fields of the FLAMINGO simulations and compare them with their gravity-only predictions. We find that, while the gas velocities do not differ from those of dark matter on large scales, the gas mass power spectrum is suppressed by up to $\approx 8\%$ relative to matter, even on gigaparsec scales. This is a consequence of star formation depleting gas in the densest and most clustered regions of the universe. On smaller scales, \textcolor{black}{$k>0.1 \, h / \rm Mpc$}, the power suppression for both gas densities and velocities is more significant and correlates with the strength of the \textcolor{black}{active galactic nucleus (AGN)} feedback. The impact of feedback can be understood in terms of outflows, identified as gas bubbles with positive radial velocities ejected from the central galaxy. With increasing feedback strength, the outflowing gas has higher velocities, and it reaches scales as large as $10$ times the virial radius of the halo, redistributing the gas and slowing its average infall velocity. Interestingly, different implementations of AGN feedback leave distinct features in these outflows in terms of their radial and angular profiles and their dependence on halo mass. In the future, such differences could be measured in observations using, for example, the kinetic Sunyaev-Zeldovich effect. }

   \maketitle
%

\section{Introduction}

In the current cosmological paradigm, baryons make up around $15\%$ of the total matter in the Universe, while the rest is in dark matter \citep{Planck:2016}. \textcolor{black}{Dark matter is collisionless and interacts only through gravity. Thus, it is hard to detect but relatively easy to model analytically on large scales or numerically in the non-linear regime. On the contrary, baryons are relatively easy to observe, but they are subject to hydrodynamical forces and a variety of astrophysical processes, which makes them difficult to model theoretically, especially on small scales where their effect is larger.} 

\textcolor{black}{The unknown mapping between gas and underlying dark matter poses strong challenges to fully extracting the cosmological information from ongoing and future large-scale structure (LSS) surveys. To constrain our cosmological models at the level of precision required, we need to take both these ingredients into account, or severe biases could appear in our cosmological analyses \citep[e.g.][]{Semboloni:2011}. For example, weak gravitational lensing (WL), which directly probes the cosmic matter field \citep{des, KIDS, HSC}, is particularly sensitive to the effects of baryons.}

\textcolor{black}{Nevertheless, since cosmic gas traces the underlying dark matter, it can be used as a powerful cosmological probe. For instance, galaxy clusters detected through the thermal Sunyaev-Zeldovich \citep[tSZ, ][]{Sunyaev:1972,Sunyaev:1980} effect or X-ray emission by their hot gas, are used to constrain the structure growth, the total neutrino mass, and the background expansion of the Universe \citep[e.g.][]{spt:2024, ghirardini:2024}. Additionally, the kinetic Sunyaev-Zeldovich \citep[kSZ, ][]{Sunyaev:1972,Sunyaev:1980} power spectrum provides information on the large-scale gas mass and velocity field of the Universe, which is sensitive to primordial non-Gaussianities \citep{Munchmeyer:2019}, neutrino mass \citep{Mueller:2015, Roncarelli:2017}, dark energy \citep{Hernandez_Monteagudo:2006, Bhattacharya:2008}, the law of gravity \citep{Bianchini:2016, Roncarelli:2018}, and the duration of reionization \citep{Zahn:2005, McQuinn:2005}. }

For all these scientific cases, it is necessary to carefully quantify the relationship between the spatial properties of gas and the underlying cosmological models, considering the impact of star formation, cooling, and feedback. This detailed knowledge is paramount for optimally exploiting the data from current and next-generation cosmological surveys, increasing the robustness of cosmological analyses, or using it directly to constrain cosmological parameters. 

Predicting the cosmic matter field in the presence of baryons is challenging. The gravitational evolution of small matter perturbations into a highly non-linear field can be accurately described by $N$-body simulations \citep[for a review, see][]{Angulo:2022}. These simulations have relatively well-understood convergence and range of validity \citep{Power:2003, Power:2006, Knebe:2009, Ludlow:2019}, and comparisons between different codes \citep{Schneider:2016} agree at the per cent level. However, baryons are subject to hydrodynamical forces and astrophysical processes that are much harder to model. 

Hydrodynamical simulations evolve dark matter and gas particles under the simultaneous effect of gravitational and hydrodynamical forces. Unfortunately, simulations featuring cosmological volumes cannot resolve galaxy formation and astrophysical processes ab initio because of the vast dynamic range of scales involved. Thus, effective subgrid prescriptions need to be adopted, for example, to model star formation, black hole seeding and growth, stellar winds, supernovae explosions, and AGN feedback, among others \citep[for a review, see][]{Vogelsberger:2020}. These astrophysical processes can impact the gas distribution up to relatively large scales. Consequently, the predictions of hydrodynamic simulations depend on the choices of the subgrid prescriptions and their effective parameters \citep[e.g.][]{VanDaalen:2011}. 
 
The subgrid parameters that regulate AGN feedback have the most significant impact on the large-scale distribution of the gas. These parameters vary the details of the energy injection from the accretion of supermassive black holes to the surrounding gas. \textcolor{black}{Massive galaxy clusters retain the universal baryon fraction within their gravitational wells, even though they are not closed systems \citep[see, for example,][]{Mitchell:2022}}. On the contrary, in galaxy groups, astrophysical processes are sufficiently energetic to eject gas from the halo, accumulating gas mass in the halo outskirts and making its density profile much more extended than that of dark matter \citep[e.g.][]{Velliscig:2014, vanDaalen:2014, Davies:2019, Tollet:2019, Angelinelli:2022, Sorini:2022, Ayromlou:2023b, Sorini:2024}. This mass redistribution has a non-negligible impact on the summary statistics of the cosmic matter field. 


The baryonic effects on the 2-point statistics of the matter field have been extensively studied with hydrodynamical simulations \citep[e.g.][]{VanDaalen:2011, Mummery:2017, Springel:2018, Chisari:2018, VanDaalen:2020, Hernandez-aguayo:2023, Schaller:2024b}. They can be modelled with flexible approaches such as the baryonification \citep{Schneider:2015, Schneider:2019, Arico:2020, Arico:2021} or the halo model \citep{Semboloni:2011, Semboloni:2013, Debackere:2020, Mead:2020}\textcolor{black}{, or models directly based on hydrodynamic simulations \citep[e.g.][]{Salcido:2023}}. The amplitude of the baryonic effect on the matter power spectrum seems to be tightly connected to the baryon fraction retained in haloes \citep{VanDaalen:2020, Salcido:2023}, even if secondary effects such as the shape of the gas density profiles need further investigations. 


Nevertheless, the analysis of the cosmic gas statistics has received comparatively little attention. On large scales, \cite{Park:2018} showed that in Illustris simulations \citep{Vogelsberger:2013, Nelson:2015}, gas is "anti-biased'' compared to the dark matter distribution. On small scales, \cite{Springel:2018} showed that in IllustrisTNG simulations \citep{IllustrisTNG}, gas clustering is suppressed, being, however, more clustered than the dark matter at higher redshifts. \textcolor{black}{\cite{Mead:2020} presented various auto- and cross-correlations of the gas field for BAHAMAS, which they used to calibrate the HMx halo model \citep{Mead:2021}}. More recently, \cite{Ni:2023} compared the gas power spectra of ASTRID \citep{astrid}, IllustrisTNG \citep{Weinberger:2017} and SIMBA \citep{Dave:2019} models in small simulations of $25 h^{-1} \rm Mpc$ box size, finding that SIMBA (ASTRID) predicts the strongest (weakest) baryonic feedback among them. \textcolor{black}{Notice that, as pointed out by \cite{Schaller:2024b}, these power spectra may not be converged due to the limited volume of the simulations, likely tending to overestimate the baryonic suppression.}

The impact of baryons on the cosmic velocity field has been studied even more seldom than gas density. \cite{Kuruvilla:2020} found that gas pairwise velocities are suppressed on small scales relative to a gravity-only universe in different hydrodynamic simulations. Strikingly, they observe that on large scales, $r \ge 10 h^{-1}\rm Mpc$, gas velocities are still suppressed by a constant factor of a few percent. Recently, \cite{Kwan:2024} did not find any indication of a velocity bias in the radial velocity profiles around haloes in the BAHAMAS simulations \citep{bahamas}. Moreover, according to \cite{Kwan:2024}, the radial velocity profiles around haloes of different masses are almost unaffected by feedback, both for the total matter and the gas. Contrary to \cite{Hellwing:2016}, who analysed the EAGLE simulation \citep{eagle}, they found that baryonic physics leaves a non-negligible imprint on the redshift-space power spectrum of the total matter field, probably due to the stronger feedback in BAHAMAS compared to EAGLE. 

This work aims to study the cosmic gas density and velocity fields relative to the case where only gravity is considered. We build on top of previous existing works and extend them leveraging the FLAMINGO cosmological-hydrodynamical suite of simulations \citep{Schaye:2023, Kugel:2023}. Due to their unprecedentedly large boxes and many subgrid-parameter variations, these simulations are well suited to study the impact of baryonic physics on cosmological scales.

Instead of focusing on any specific observable, we study the impact of galaxy formation on several statistics of the gas field and identify the physical processes that drive those differences. We seek to build a physical picture of the relevant baryonic processes that should be considered when modelling the cosmic gas density and velocity fields. We will demonstrate that 1) on large scales, \textcolor{black}{$k<0.1 \,  h / \rm Mpc$,} gas moves as dark matter, but it is less clustered than dark matter \textcolor{black}{due to the most highly biased gas being converted into stars}; 2) On small and intermediate scales, \textcolor{black}{$k>0.1 \, h /\rm Mpc$ or $r < 10 \, h^{-1} \rm Mpc$,} baryonic effects suppress both gas clustering and velocities; 3) Feedback changes the gas density and velocity profiles up to the outskirts of group-size haloes; 4) The outflowing gas reaches several virial radii, and its properties depend on feedback strength as well as feedback implementation scheme. 

The paper is organised as follows. We present the FLAMINGO simulations in Section \ref{sec:simulations}. In Section \ref{sec:field}, we study the cosmic gas density and velocity fields and characterise the main effects that drive the differences with respect to the gravity-only case. In Section \ref{sec:haloes}, we focus on the astrophysical processes that intervene in haloes of different mass, disentangling features that arise from different subgrid parameters. Finally, in Section \ref{sec:conclusions}, we summarise the main results and conclude. 

\section{Simulations}\label{sec:simulations}

This work is based on the state-of-the-art FLAMINGO cosmological-hydrodynamical simulations \citep{Schaye:2023}. The suite features different box sizes and resolutions. Here, we employ the intermediate resolution, that is $N=1800^3$ dark matter and baryonic particles, and $N=1000^3$ neutrino particles in boxes of $L=1000 \rm \, Mpc = 681 h^{-1} \rm Mpc$. The average gas particle mass is $m_{\rm g}=1.07 \cdot 10^9  h^{-1} \rm M_{\odot}$, while the dark matter particle mass is $5.65 \cdot  10^9  h^{-1} \rm M_{\odot}$ in the full-physics run and $6.72 \cdot 10^9  h^{-1} \rm M_{\odot}$ in the gravity-only run. Unless stated otherwise, all the results in this paper are at the intermediate resolution and fiducial cosmology at $z=0$. Details of the initial conditions, neutrino implementation, hydrodynamic solver and subgrid prescriptions are given in \cite{Schaye:2023}. \textcolor{black}{In the following, we briefly discuss the subgrid models for AGN feedback due to their relevance for this work.}

\textcolor{black}{The FLAMINGO suite includes two different subgrid models for AGN feedback: thermal and isotropic energy injection \citep{Booth:2009}, and kinetic or jet-like energy injection \citep{Husko:2022, Schaye:2023}. Most of the simulations adopt the thermal implementation. While the black hole (BH) is accreting gas, a fraction $\epsilon = 0.015$ of the accreted rest mass energy is added to the BH energy reservoir. Once the BH has collected enough energy to increase the temperature of a gas particle by $\Delta T_{\rm AGN}$, the thermal energy is injected into the closest SPH particle to the black hole. Statistically, this becomes an isotropic energy injection. The main parameter controlling the strength of AGN feedback is $\Delta T_{\rm AGN}$, which is left free in the calibration.}

\textcolor{black}{The kinetic or jet AGN model implemented in FLAMINGO is a simplification of the subgrid model presented in \cite{Husko:2022}. The same energy as in the thermal implementation is available to the BH, the only difference being how the energy is injected back into the surrounding gas. Once the energy reservoir of the BH exceeds $2 \times \frac{1}{2}\times m_g \times v_{\rm jet}^{2}$, two gas particles are randomly selected in a cone around the BH spin direction (each on one side), and kicked with velocity $v_{\rm jet}$. Analogous to the $\Delta T_{\rm AGN}$ in the thermal model, $v_{\rm jet}$, the velocity of the jet controls the AGN feedback strength, and its value is set in the calibration.}


A powerful aspect of the suite is that it contains several astrophysics variations -- choices for the subgrid parameters -- run with the same code, resolution, box size and cosmology, so the impact of astrophysics can be isolated. The fiducial parameter set has been calibrated to agree with the measured gas mass fractions in low-redshift galaxy clusters and the galaxy stellar mass functions at $z=0$ \citep{Kugel:2023}. Other variations are calibrated to reproduce deviations from observed gas fractions in clusters (for example, $\text{fgas}+2 \sigma $ and $\text{fgas} - 8\sigma $) or from the observed stellar mass functions (for example $M_* - \sigma $). Notice that the calibrations are not done "by hand``, but using Gaussian process emulators \citep[the calibration procedure is explained in detail in][]{Kugel:2023}. Four subgrid parameters, two controlling stellar feedback and two controlling AGN feedback, are fitted in the calibration; however, the parameters that control the gas fractions in clusters are mainly related to the AGN feedback efficiency. The parameters that control the stellar mass function are instead primarily related to supernova feedback \citep[see Table 1 of ][]{Schaye:2023}. The suite also contains a "no cooling`` run that was first presented in \cite{McCarthy:2024}. This run does not form stars and, consequently, does not feature feedback, which helps disentangle the effects of galaxy formation from other hydrodynamical processes. Moreover, the simulations with different AGN feedback prescription \textcolor{black}{(but calibrated to the same observables)}, are beneficial to test the impact of the adopted subgrid model and its numerical implementation on the final observables. 

Structure finding is done using the VELOCI\textsc{raptor} code \citep{Elahi:2019}, which builds on top of the Friends-of-Friends groups \citep[FoF,][]{Press:1982, Einasto:1984, Davis:1985}, finding dynamically distinct groups performing an iterative 6D FOF search in the phase space of the original FoF groups. The centre of the haloes is the position of the most bound particle. Then, Spherical Overdensity and Aperture Processor (SOAP) is run on top of the previously identified haloes and subhaloes to obtain 3D or projected properties, among other options \citep{Schaye:2023}. In this work, we employ the spherical overdensity outputs (SO) with enclosed density $\Delta = 200 \rho_{\rm m}$, where $\rho_{\rm m}$ is the mean density of the universe. All the results shown are for isolated main haloes without considering satellite subhaloes.  

\section{Gas statistics: mass distribution and velocity}\label{sec:field}

In this section, we characterise the gas mass and velocity fields in the FLAMINGO suite. We study how non-gravitational physics impacts several summary statistics on different scales. In particular, we investigate the density power spectrum (Section \ref{sec:dens_ps}), velocity divergence power spectrum (Section \ref{sec:vdiv_ps}), and mean pairwise velocities (Section \ref{sec:pairwise}). 

\subsection{Density power spectrum}\label{sec:dens_ps}

\begin{figure*}
    \centering
    \includegraphics[width=\linewidth]{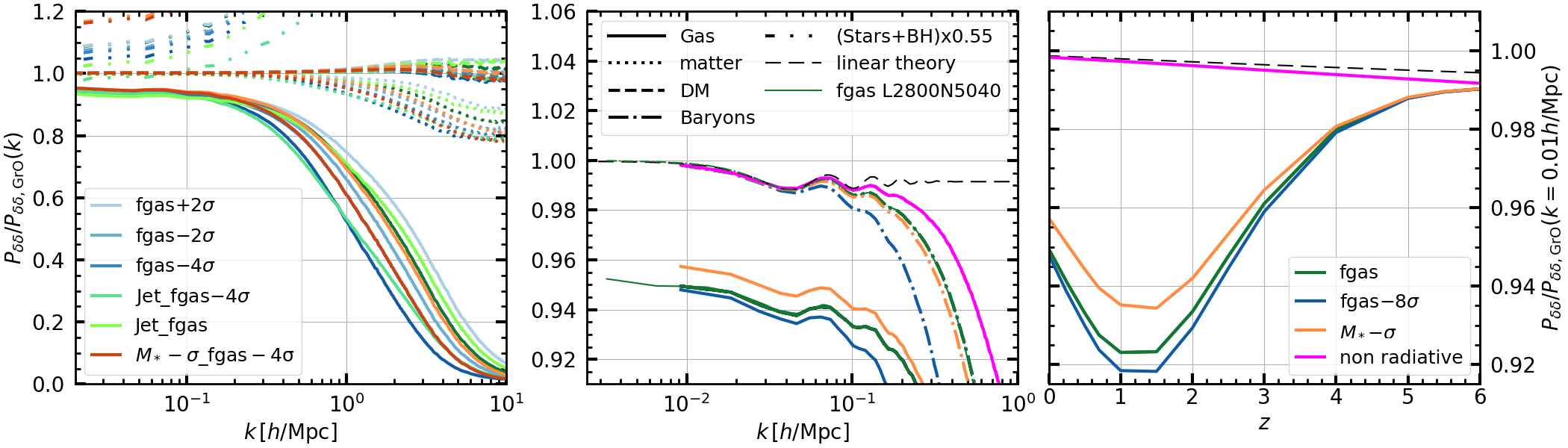}
    \caption{Ratio of density power spectra between hydrodynamic and gravity-only runs for several species. The colours indicate different FLAMINGO models. For comparison, we add the non-radiative run. \textit{Left panel}: results at $z=0$ for gas (solid lines), dark matter (dashed lines), stars and black holes (dashed double-dotted lines), and total matter (dotted lines). \textit{Middle panel}: Zoom of the large-scale suppression of four representative FLAMINGO models in the gas (solid lines) and total baryons (dashed-dotted lines) components at $z=0$. \textit{Right panel}: Redshift evolution of the large-scale gas bias, defined as the ratio between gas and gravity-only power spectra at $k=0.01 h \rm /Mpc$ at each redshift.}
    \label{fig:power-spectra}
\end{figure*}

The impact of baryonic processes on the distribution of a given component is encoded in the ratio of its power spectrum with the gravity-only power spectrum, $S_{\delta \delta}(k) = P_{\delta \delta} / P_{\delta \delta, \rm GrO} (k)$. The power spectrum is defined as $P_{\delta \delta, i} (k) =\langle |\hat{\delta_i}(k)|^2\rangle $, where $\hat{\delta_i}$ is the Fourier transform of $\delta_i = (\rho_i - \bar{\rho}_i) / \bar{\rho}_i$, and $\rho_i$ and $\bar{\rho}_i$ are the density and mean density of the corresponding species, respectively. If the ratio $P_{\delta\delta, i} / P_{\delta\delta,\rm GrO}$ is 1, the clustering is identical to the gravity-only run. All power spectra are shot-noise corrected, following equation B.11 from \cite{Mead:2020}. In the left panel of Figure \ref{fig:power-spectra}, we plot the results for different species: gas, stars and black holes, dark matter, and total matter (gas + dark matter + stars + black holes, dotted lines). The different colours denote the different FLAMINGO simulations. 

The two main baryonic processes that shape the distribution of matter with respect to a purely gravitational universe are i) the formation of massive galaxies in the centre of haloes and ii) AGN feedback, which pushes gas outside haloes. Consequently, the dark matter relaxes according to the modified matter distribution and gravitational field. The relevance of these baryonic processes varies depending on the matter species and scales we are looking at. 

Stars and black holes (gas) are the most (least) clustered species on small scales. The strength of the AGN feedback mainly drives the clustering suppression of gas over the gravity-only run. For models calibrated to low gas fractions, such as $\text{fgas}-8\sigma$, the gas clustering is suppressed by $95\%$ at $k=3 \, h \rm Mpc^{-1}$, while models with higher gas fractions, such as $\text{fgas} +2\sigma$, attain the same suppression only on smaller scales, $k > 10 \, h \rm Mpc^{-1}$. By $k=7 \cdot 10^{-1} h \rm Mpc^{-1}$, gas clustering is suppressed by more than $20 \%$ in all models considered. 

The dark matter back-reacts to the combined effect of all the baryonic components. Its clustering differs from a gravity-only universe by $P_{\rm dm} \approx 1.05-0.95 P_{\rm GrO}$ on scales $k<10 \, h \rm Mpc^{-1}$, being affected by both the creation of stars (enhancing the clustering) and the expulsion of gas (lowering the clustering). As gas is much more abundant than stars and black holes, the total matter clustering is suppressed between $P_{\rm matter} \approx 0.8-1.0 P_{\rm GrO}$.

\subsubsection{Large scale bias}\label{sec:dens_ps_large}




On large scales, the dark matter and total matter clustering in the full-hydro run trace the gravity-only run. However, on the same scales ($k \le 2 \cdot 10^{-2} h \rm Mpc^{-1}$) the gas is biased low by a constant $4-5\%$, irrespective of the specific AGN feedback. Similarly, on the same scales, stars and black holes are biased high by $1.8-2.2$. This is unrelated to the different masses of the gravity-only, gas, stellar, and black hole fields. As explained at the beginning of this section, the power spectra are normalised so that their ratios go to $1$ when they have the same clustering, irrespective of the total mass of each species. 

In the middle panel of Figure \ref{fig:power-spectra}, we look more closely at the large-scale gas bias. We have plotted the suppression for four relevant FLAMINGO models: the fiducial model (solid blue, $\text{fgas}$), one with extremely low gas fractions in clusters (solid purple, $\text{fgas} -8\sigma$), and one with fewer stars (solid brown, $M_* - \sigma$). We also add the non-radiative run, which does not form galaxies and, therefore, has no feedback (solid magenta). For the fiducial model, we have added the suppression from a larger box with $L=2800 \, \rm Mpc$ but the same resolution (thin green line): on scales $k=10^{-2} h \rm Mpc^{-1}$ the suppression is converged. 

We observe that FLAMINGO models sharing the same stellar mass function converge to the same suppression of $\approx 5\%$, with sub-percent differences irrespective of the gas fractions ($\text{fgas}$ vs $\text{fgas} - 8 \sigma$). The gas bias is slightly weaker in models calibrated to a reduced stellar mass function, reaching $4\%$ ($M_* - \sigma$). These differences disappear in the non-radiative run, or once we take into account all baryons (dashed-dotted lines): baryons perfectly trace the gravity-only clustering at $k<10^{-2} h \rm Mpc^{-1}$. On smaller scales, the differences between baryons' and dark matter's primordial distributions are not completely washed out yet by gravity \citep{Angulo:2013} and a residual suppression of $~1 \%$ remains, consistent with the linear prediction from CLASS at $k<10^{-1} h \rm Mpc^{-1}$ (black dashed line). 

We interpret the large-scale gas bias as a necessary consequence of the large-scale stellar bias. As gas turns into stars in high-density regions, presumably haloes — which are more clustered and biased compared to the dark matter distribution — the stars preferentially form and live in more clustered environments relative to the dark matter field (with bias larger than one). Conversely, gas mass is reduced in those high-density environments, or, in other words, it preferentially lives in lower-density and less clustered environments (with a bias of less than one). In agreement with this interpretation, the large-scale gas (stellar) bias is also smaller (larger) in the model calibrated to a reduced stellar mass function, and there is no gas bias in the non-radiative run. 

In the right panel of Figure \ref{fig:power-spectra}, we plot the redshift evolution of the large-scale gas bias, defined by the power-spectrum suppression at $k=0.01 h/\rm Mpc$ at each redshift. The non-radiative run closely follows the linear theory prediction for the total baryons. Until $z \approx 6$, the gas distribution agrees with it, but once stars start forming, the gas distribution becomes biased. In all redshifts, the bias in the $M_*-\sigma$ run is lower compared to the fiducial run. The discrepancy between $\text{fgas}$ and $\text{fgas}-8\sigma$ is larger at higher redshift, which could be due to their stellar mass functions being different at higher redshift (note that the calibration was done to reproduce the same stellar mass functions at $z=0$, which does not guarantee the same at higher redshifts).

Interestingly, gas is most biased in all models at $z \approx 1$, becoming less biased at lower redshifts. We speculate that this could be again a consequence of the star formation history, the stellar bias becoming smaller at lower redshifts \citep[e.g.][]{Fry:1996, Tegmark:1998, Springel:2018}, simultaneously reducing the bias in the gas field. We explore this more in Appendix \ref{sec:gas_bias_z}. 

Earlier studies by \cite{Shaw:2012} and \cite{Park:2018} identified a significant gas bias in other hydrodynamic simulations, including Illustris \citep{Vogelsberger:2013, Nelson:2015}. We confirm their findings and extend them to the larger scales explored in the FLAMINGO simulations. Although the redshift evolution and the amplitude of the bias vary quantitatively, the qualitative trends and their interpretation remain consistent.

Thus, care is required when interpreting observations that employ gas as a tracer. For instance, \textcolor{black}{in halo model approaches,} the gas bias should be considered on top of the halo bias when analysing the thermal Sunyaev-Zeldovich (tSZ) power spectrum measurements \citep{Bolliet:2018, Salvati:2018, Douspis:2022}, kSZ power-spectrum measurements \citep{Shaw:2012, Park:2018}, or halo bias-weighted mean electron pressure of the universe \citep{Chiang:2020, Chiang:2021, Chen:2023, Chen:2024a}. \textcolor{black}{The analyses based on hydrodynamic simulations inherently account for the gas bias \citep[e.g.][]{Battaglia:2012,Battaglia:2012b,McCarthy:2014}.}

\subsection{Velocity-divergence power spectrum}\label{sec:vdiv_ps}

\begin{figure}
    \centering
    \includegraphics[width=0.8\linewidth]{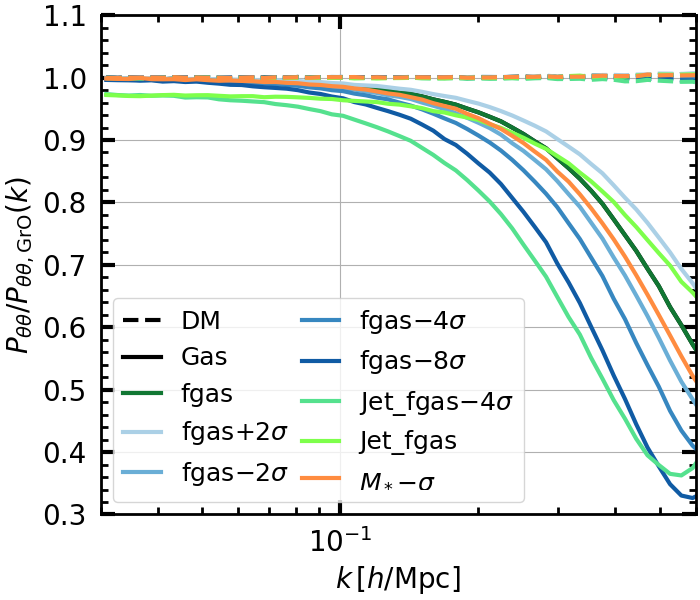}
    \caption{Ratio of velocity-divergence power spectra between hydrodynamic and gravity-only runs for several species at $z=0$. The different colours indicate different FLAMINGO models. We display the results for gas (solid lines) and dark matter (dashed lines).}
    \label{fig:power-spectra-vdiv}
\end{figure}

In linear theory, the peculiar velocity field follows a pure potential flow (i.e. without any rotational component), fully described by
\begin{equation}\label{eq:vel_dens_lin_0}
    \theta = \nabla \cdot v = - \delta afH,
\end{equation}
where $\delta$ is the density contrast, $a$ is the expansion factor, $f=\frac{d \ln D(a)}{d \ln a}$ is the linear growth rate and $H$ is the Hubble parameter. Thus, given the density field, the velocities are known: particles move convergently towards high-density regions and divergently away from low-density regions. However, the universe is no longer linear at low redshifts, and the velocity field becomes uncorrelated with the density field as we approach small scales. Gravitational collapse and several baryonic processes contribute to this decorrelation. In this section, we focus on the latter.

We compute the velocity-divergence using the Delaunay-tessellation field interpolator (DTFE) technique \citep{Schaap:2000, Cautun:2011}. The DTFE package performs a Delaunay triangulation of particles and then interpolates the fields to a regular grid. Even though it has some limitations, velocity-divergence power spectra are well-behaved \citep{Pueblas:2009, Hahn:2015}. However, as the velocity-divergence is a coarse-grained quantity, as we increase the resolution of the mesh, the value of the velocity-divergence can change dramatically. We compute the velocity-divergence and its power spectrum using a mesh of $N=256^3$ cells. By comparing the power spectra with higher and lower resolution meshes, we find that it is converged only at $k < 6 \cdot 10^{-1} h / \rm Mpc$, which defines the smallest scale we investigate. 

In Figure \ref{fig:power-spectra-vdiv}, we plot the ratio of the velocity-divergence power spectra of hydrodynamic and gravity-only simulations, $S_{\theta\theta}(k)$. The power in potential flows of gas (solid lines) is suppressed compared to that of dark matter, both for the dark matter in the full-physics run and in the gravity-only run (dashed lines). The dependence on the baryonic model is similar to for $S_{\delta \delta}$. It is interesting to note that, while the gravitational non-linearities act oppositely on densities and velocities (increasing the mass clustering and suppressing potential flows) \citep{Pueblas:2009, Zhang:2013, Zheng:2013}, the baryonic effects act in the same direction, suppressing both the clustering and potential flows.

Furthermore, unlike density clustering, velocity divergence is not biased on large scales in any FLAMINGO model \footnote{$\text{Jet\_fgas}$ and $\text{Jet\_fgas}-4\sigma$ exhibit a small $\approx 2\%$ bias on large scales. We attribute this to an unexplained behaviour of jet simulations, which causes the power spectra to not converge at large scales rather than to a physical signal. In other works, such as \cite{Schaller:2024b}, this disagreement was corrected in post-processing. We decided to display the raw data, as it is not evident how to correct it in some of the quantities we show in the paper, such as the velocity divergence power spectra or the pairwise velocities.}. Notice that this aligns with the expectations from linear theory (Eq. \ref{eq:vel_dens_lin_0}): the gas feels the gravitational force from the entire matter field (not just the gas), which remains unbiased on large scales. Finally, the scale-dependent suppression of power starts on larger scales for the velocity divergence than for the density: for instance, for $\text{fgas} - 8\sigma$ at $k=4 h / \rm Mpc$, $S_{\delta \delta} \approx 0.8$, while $S_{\theta \theta} \approx 0.6$. Therefore, even if a fraction of the suppression of power in velocity-divergence can be understood as a result of different mass distributions in the full-physics and gravity-only runs (via Equation \ref{eq:vel_dens_lin_0}), it is clear that some baryonic processes further destroy the potential flows of gas and are efficient on very large scales. 

\subsubsection{Correlation between density and velocity divergence}

\begin{figure}
    \centering
    \includegraphics[width=0.8\linewidth]{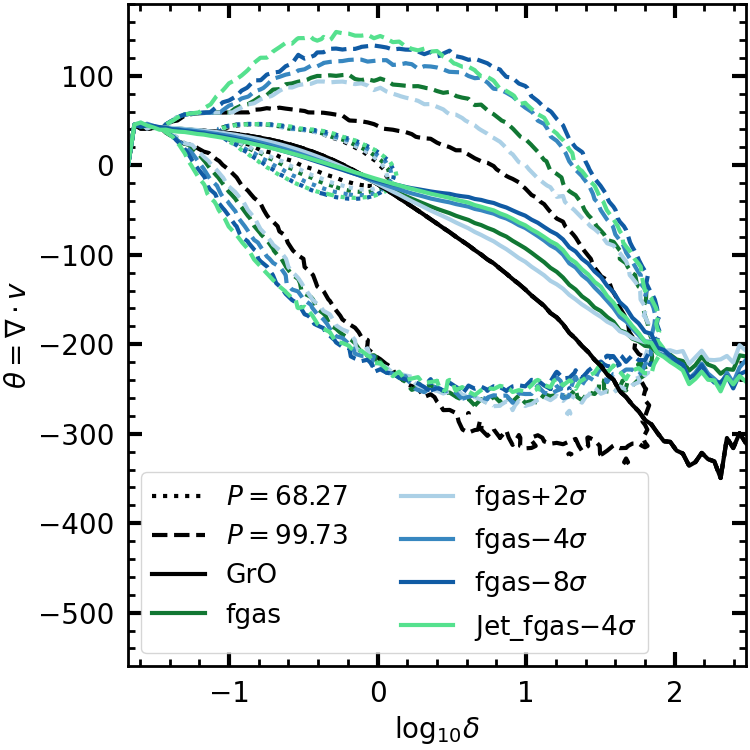}
    \caption{Contours showing the correlation between velocity divergence and density contrast in the gravity-only (black lines) and gas in the hydrodynamic runs (coloured lines). Different FLAMINGO models are represented with different colours. The solid lines show the mean of the distribution, while the dotted and dashed lines define the $68.3$ and $99.7$ percentiles of the distribution, respectively.}
    \label{fig:vdiv_dens}
\end{figure}

Figure \ref{fig:vdiv_dens} illustrates the correlation between velocity-divergence and density for the gravity-only simulation (black lines) and gas in several FLAMINGO models (coloured lines). The density and velocity divergence are computed on a mesh of $N=256^3$ cells, with a cell size of $2.66 h^{-1} \rm Mpc$ using the Delaunay-tesselation field interpolator \citep{Cautun:2011}. The dark matter in the full-physics runs yields similar results to the gravity-only run, so we do not plot it for clarity. The dotted and dashed lines represent the $68.3$ and $99.7$ percentiles of 2d histograms, while the solid line shows the mean of the relation.

At low densities, $\log_{10} \delta < -1$, the relation is tight for gas and dark matter, and the mean value is very similar -- linear theory still holds. As we go to higher-density regions, the scatter around the mean increases, and by $\log_{10} \delta > 0$, the mean starts to differ between dark matter and gas. High-density regions become shell-crossed, potential flows are destroyed, and transversal motions --i.e., the velocity field curl component-- arise \citep{Zhang:2013, Zheng:2013, Hahn:2015}. This purely gravitational effect contributes to the scatter in the dark matter and gas. Yet, baryonic processes affect the gas on top of it. On the one hand, there is a clear distinction in the high-density tails of the distribution, visible in the $P=99.7$ percentiles: the convergent motion into high-density regions stops faster for gas than in dark matter (at $\theta_{\rm gas} \approx -250$ vs $\theta_{\rm DM} \approx -300$), almost independently of the model. This is consistent with the explanation that gas gets shocked when infalling into high-density environments, and its kinetic energy is converted into heat, contributing to the decrease of convergent motions. On the other hand, the dependence on the AGN strength appears in the divergent tail of the distribution ($\theta > 0$) at all densities. We interpret this as gas ejected by the AGN, which has positive radial velocities, i.e. directed away from the high-density regions.

\subsection{Mean pairwise velocity}\label{sec:pairwise}

\begin{figure}
    \centering
    \includegraphics[width=\columnwidth]{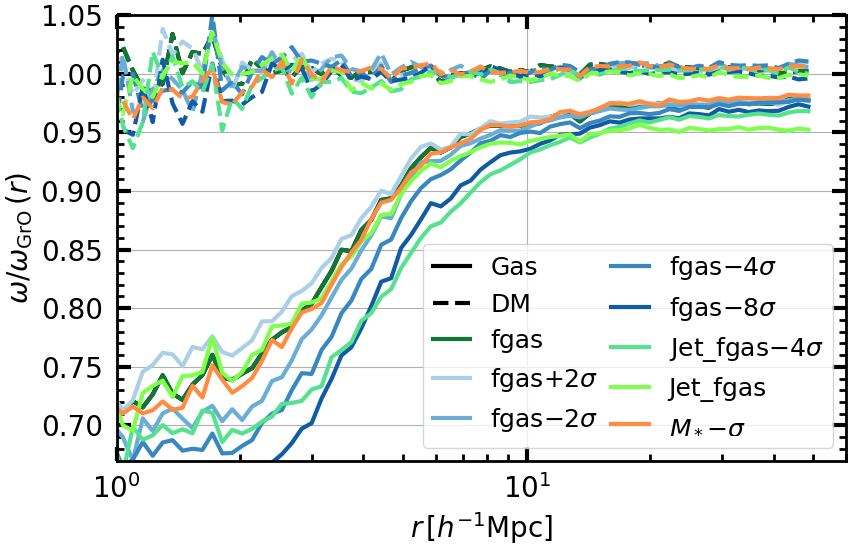}
    \caption{Ratio of mean pairwise velocities between hydrodynamic and gravity-only runs for the gaseous (solid lines) and dark matter (dashed lines) components at $z=0$. The colours represent different FLAMINGO models.}
    \label{fig:pairwise-velocities}
\end{figure}

Another common way to study the velocity field is using pairwise velocities. Pairwise velocities tell us how two particles are moving with respect to each other as a function of the separation scale. Even though the distribution is broad (\cite{Kuruvilla:2020}, figure 1), on average, particles are getting closer to each other due to the attractive nature of gravity. In other words, the mean pairwise velocity $\omega (r)  $ is always negative, 
\begin{equation}\label{eq:pairw}
     \omega (r)  = \frac{\langle (1+\delta_1)(1+\delta_2)(\vec{v}_2 - \vec{v}_1) \cdot \hat{r}\rangle }{\langle (1+\delta_1)(1+\delta_2)\rangle } < 0,
\end{equation}
where $\vec{r} = \vec{r}_2-\vec{r}_1$, $\delta_i = \delta(\vec{r}_i)$, $\vec{v}_i = \vec{v}(\vec{r}_i)$ and the average is over all particle pairs with separation $r$. Notice that this is a mass-weighted quantity because we sample the velocity field only where particles are located. At leading order, Eq. \ref{eq:pairw} reduces to (a full derivation is given in Appendix \ref{sec:pairwise_derivation})
\begin{equation}\label{eq:pairw_lead} 
      \omega(r)   = [\langle \delta_1\vec{v}_2\rangle  - \langle \delta_2\vec{v}_1\rangle ] \cdot \hat{r}.
\end{equation}
This expression shows we are, in fact, dealing with the \textit{pairwise momentum} rather than pairwise velocity. As defined here, both the density and velocity fields contribute to the pairwise velocities.

In Figure \ref{fig:pairwise-velocities}, we plot the ratio of the mean pairwise velocity in hydrodynamic and gravity-only simulations, $R_{\omega}= \omega(r) /  \omega_{\rm GrO}(r) $. The mean pairwise velocities were computed using the \textsc{halotools} package \citep{halotools} with some modifications, with $100^3$ randomly selected particles. 

Gas pairwise velocities are suppressed on all scales, separated into two regimes. On small scales, $r<10 h^{-1} \rm Mpc$, pairwise velocities are slower in a model-dependent way, reaching to $25-30\%$ suppression at $r=1 h^{-1} \rm Mpc$. The models with stronger AGN feedback predict stronger suppression. At large scales, $r>10 h^{-1} \rm Mpc$, all models converge towards a constant bias $\approx 2-3\%$, though some model dependency remains. 

Similar results were found in other hydrodynamic simulations by \cite{Kuruvilla:2020}. There, they argued that the reason for the large-scale bias was that stellar feedback prevents gas from infalling. However, the interpretation of pairwise velocities in general and large-scale "velocity bias" in particular is not straightforward. As Eq. \ref{eq:pairw_lead} shows, one cannot tell whether gas is infalling slower than dark matter or if the measured bias is an effect of the mass-weighted nature of the velocity field. 

\begin{figure}
    \centering
    \includegraphics[width=0.9\columnwidth]{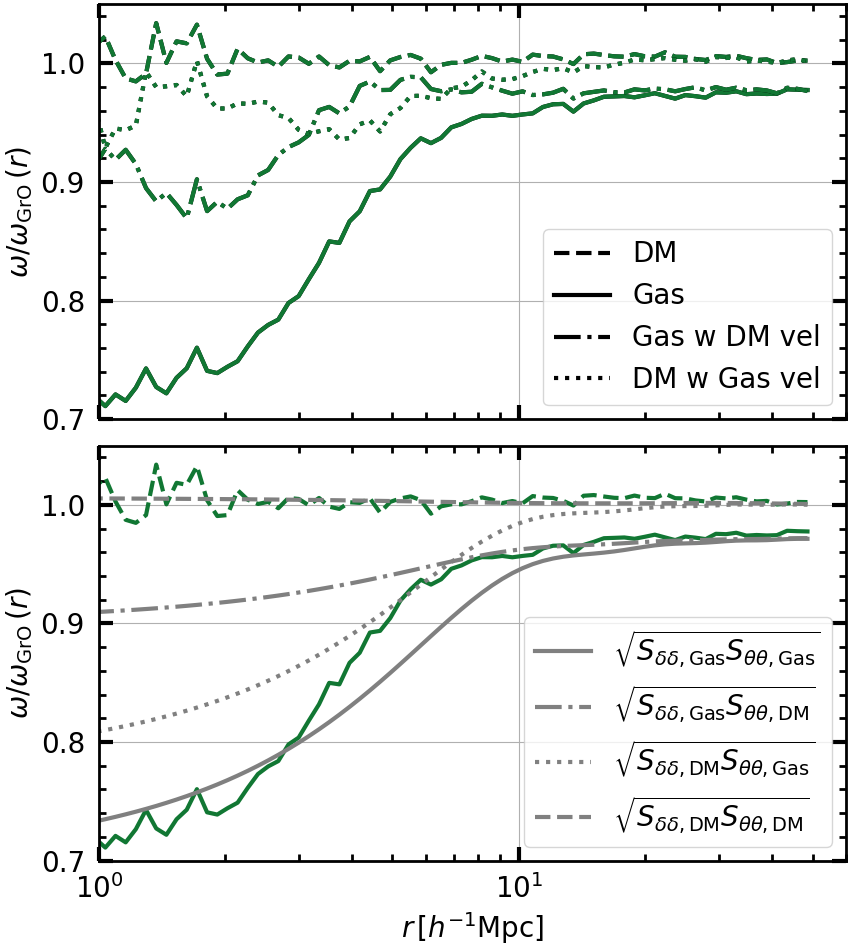}
    \caption{Ratio of mean pairwise velocities between fiducial hydrodynamic and gravity-only runs at $z=0$. The gas and dark matter ratios are displayed in both panels in solid and dashed green lines, respectively. \textit{Upper panel}: Ratio of mean pairwise velocities with dark matter velocity field sampled at gas positions (dashed-dotted lines) and gas velocity field sampled at dark matter positions (dotted lines). \textit{Lower panel}: Predictions from linear theory assuming different boosts for velocity and density power spectra, as indicated by the legend. The details of the procedure are explained in the main text.}
    
    \label{fig:pairwise-velocities-cross}
\end{figure}

We disentangle these two contributions in the upper panel of Figure \ref{fig:pairwise-velocities-cross}. Our approach consists of keeping the gas distribution but assigning the velocity field of dark matter. Effectively, per gas particle in the simulation, we find the closest dark matter particle and assign its velocity to the original gas particle\footnote{Notice that this approach works because dark matter velocities are practically the same in full physics and gravity-only runs, that is, their response to baryonic processes is negligible on the scales we are looking at.}. This can be understood as sampling the dark matter velocity field in gas positions. Dashed-dotted lines show the resulting pairwise velocity. Similarly, the result for dark matter distribution with gas velocities is shown using dotted lines.

On intermediate and small scales, $r<10 h^{-1} \rm Mpc$, both gas distribution (dashed-dotted lines) and gas velocities (dotted lines) differ from those of dark matter, suppressing the mean pairwise velocity of the gas. On these scales, gas particles indeed fall slower than dark matter. This is probably due to many effects, such as losing kinetic energy through shocks and AGN feedback. 

On large scales, $r>10 h^{-1} \rm Mpc$, the bias of gas pairwise velocities is dominated by the gas distribution: using dark matter or gas velocities (solid vs dashed-dotted lines) makes little difference. Conversely, there is a slight bias if one assumes the dark matter distribution, no matter the velocity field (dashed vs dotted lines). This means that, at large separations, gas particles infall towards each other at the same velocity as the dark matter. The bias in gas pairwise velocities mostly comes from weighting the (unbiased) velocity field with the (biased) density field. 

This interpretation is consistent with \cite{Kwan:2024} and the analysis presented in Section \ref{sec:profiles}, where we do not find any significant velocity bias, both in the FLAMINGO and BAHAMAS simulations. 

\subsubsection{Mean pairwise velocities in linear theory}\label{sec:pairwise_linth}

Using the insights from linear theory, we can make more quantitative statements about the relation between pairwise velocities and density and velocity fields. It can be shown that in linear theory (see Appendix \ref{sec:pairwise_derivation} for the full derivation),
\begin{equation}\label{eq:paiw_lin}
      \omega_l(r)   \propto \int{kP_{\delta \delta, \rm l}(k)j_1(kr)dk},
\end{equation}
where $P_{\delta \delta, \rm l}(k)$ is the linear matter power spectrum and $j_1$ is the Bessel function of the first kind. Taking into account that
\begin{equation}\label{eq:vel_dens_lin}
    -i \vec{k} \cdot \tilde{\vec{v}} = \tilde{\delta} a f H,
\end{equation}
the following relation exists between velocity, velocity-divergence and matter power spectra:
\begin{equation}\label{eq:Pvel_Pdens}
    P_{vv, \rm l}(k) = P_{\theta \theta, \rm l}(k) k^2 = P_{\delta \delta, \rm l}(k) k^2 (afH)^2.
\end{equation}
Thus, we can re-write Eq. \ref{eq:paiw_lin} as
\begin{equation}\label{eq:paiw_lin_2}
      \omega(r)   \propto \int{k \sqrt{P_{\delta \delta, \rm l} P_{\theta \theta, \rm l}} j_1(kr)dk}  \propto \int{\sqrt{P_{\delta \delta, \rm l} P_{vv, \rm l} } j_1(kr)dk}.
\end{equation} 
As in Eq. \ref{eq:pairw_lead}, the dependence on density and velocity is explicit.

These expressions are strictly valid only in linear theory and for a collisionless fluid. However, we want to apply them to gas, which is subject to other forces and phenomena, such as pressure or feedback, on top of purely gravitational forces. To that end, we modify the linear-theory expression, introducing the "baryonic suppression" $S(k)$,
\begin{equation}\label{eq:boost}
    S(k) = \sqrt{S_{\delta\delta}(k)S_{vv}(k)} = \sqrt{\frac{P_{\delta \delta}}{P_{\delta \delta, \rm GrO}}} \sqrt{\frac{P_{vv}}{P_{vv, \rm GrO}}},
\end{equation}

\noindent where now all power spectra are non-linear. Notice that with Eq. \ref{eq:Pvel_Pdens}, $S_{v v} = S_{\theta\theta}$. Because we multiply the linear theory power spectra with the ratio between non-linear power spectra, the modification is of order $1$, albeit an ad-hoc term. Finally, the pairwise velocities 

\begin{equation}\label{eq:paiw_lin_boost}
   \omega(r)  \propto \int{k P_{\delta \delta,l}(k) S(k)j_1(kr)dk}.
\end{equation}

We can directly test the impact of density and velocity fields in the pairwise velocity, changing the suppression factor in Eq. \ref{eq:boost} via the $S_{\delta \delta}$ and $S_{\theta \theta}$ measured in simulations (Figure \ref{fig:power-spectra}). The predictions for different combinations are shown in the lower panels of Figure \ref{fig:pairwise-velocities-cross} in grey. Assuming the gas mass distribution and velocities yields a fairly good description of pairwise velocities on all scales (solid line). Replacing gas velocities with dark matter velocities (dashed-dotted line) makes the suppression smaller on small scales. Thus, density and velocity fields impact the pairwise velocities on small scales by $10-20\%$. Eq. \ref{eq:paiw_lin_2} is based on linear theory and expected to fail on small scales. Hence, the small-scale behaviour should be interpreted qualitatively. 

The large-scale behaviour is set by the choice of $S_{\delta \delta}$: assuming the gas distribution yields a $\approx2.5 -3 \%$ bias (solid and dashed-dotted lines). In contrast, the dark matter distribution does not give a significant bias (dashed and dotted lines), independent of whether a gas or dark matter velocity field is used. Hence, the large-scale bias in the pairwise velocities can be explained by the large-scale bias in the gas power spectrum (Figure \ref{fig:power-spectra}).

The qualitative interpretation of such a bias (see Sec. \ref{sec:dens_ps_large}) is consistent with this bias showing up in pairwise velocities too. Haloes are clustered, and clustered regions fall towards each other faster than unclustered regions \citep{Tinker:2007, Shirasaki:2021}. According to our interpretation of the large-scale bias in the gas power spectra, because gas forms stars in the centres of haloes, it preferentially resides in lower-density regions, creating a bias $b<1$. In pairwise velocities, this implies that, relative to dark matter, the gas field weights haloes or clustered regions less, that is, it weights less the regions with the largest (or most negative) pairwise-velocities, yielding on average lower (less negative) pairwise-velocities on large scales. 

Although pairwise kSZ observations do not directly measure the mean pairwise velocity field of gas as computed here, but instead focus on halo positions \citep{Hand:2012, Calafut:2021, Chen:2022, Gong:2024, Li:2024, Schiappucci:2024}, the gas bias may still introduce a notable effect. Following the same argument as in Section \ref{sec:dens_ps_large}, one should consider that gas underweights high-density regions and introduces a gas bias on top of the halo bias to interpret the measurements accurately. Nevertheless, in the case of pairwise kSZ measurements, the overall impact is expected to be smaller since it scales with the square root of the baryonic suppression (Equation \ref{eq:boost}).





\section{Gas density and velocity in halo outskirts}\label{sec:haloes}





In the previous section, we analysed summary statistics of gas density and velocity fields and found that several astrophysical processes significantly impact them. Notably, feedback shapes both fields on small scales. In this section, we study the source of those effects: haloes. We restrict the analysis to the FLAMINGO runs calibrated to varying gas fractions within clusters and exclude the variations of the stellar mass function.

\begin{figure*}
    \centering
    \begin{subfigure}[t]{0.49\textwidth}
        \centering
        \caption{Dark matter.}
        \includegraphics[width=\textwidth]{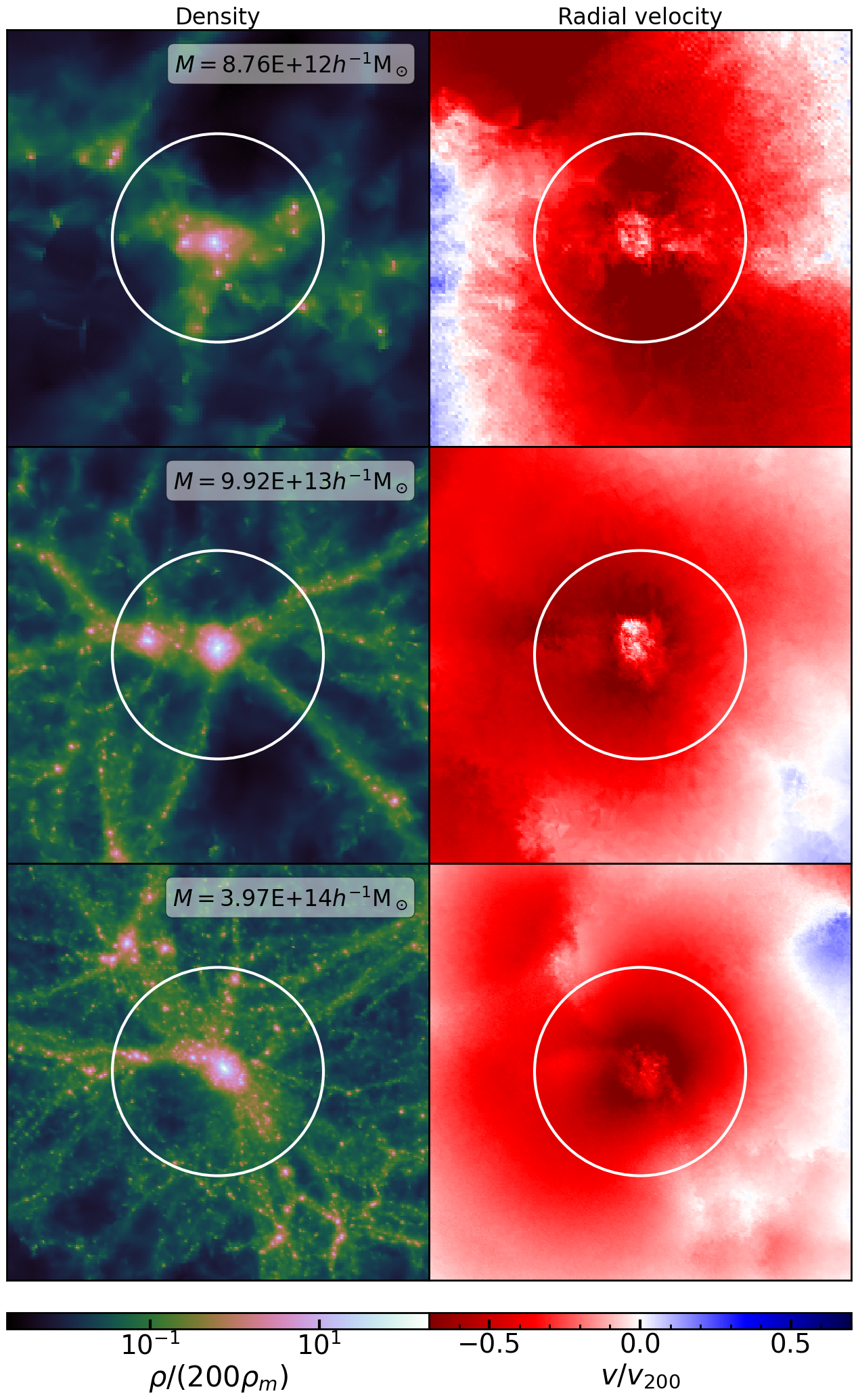}
        \label{fig:halo_example_dm}
    \end{subfigure}
    \hfill
    \begin{subfigure}[t]{0.49\textwidth}
        \centering
        \caption{Gas.}
        \includegraphics[width=\textwidth]{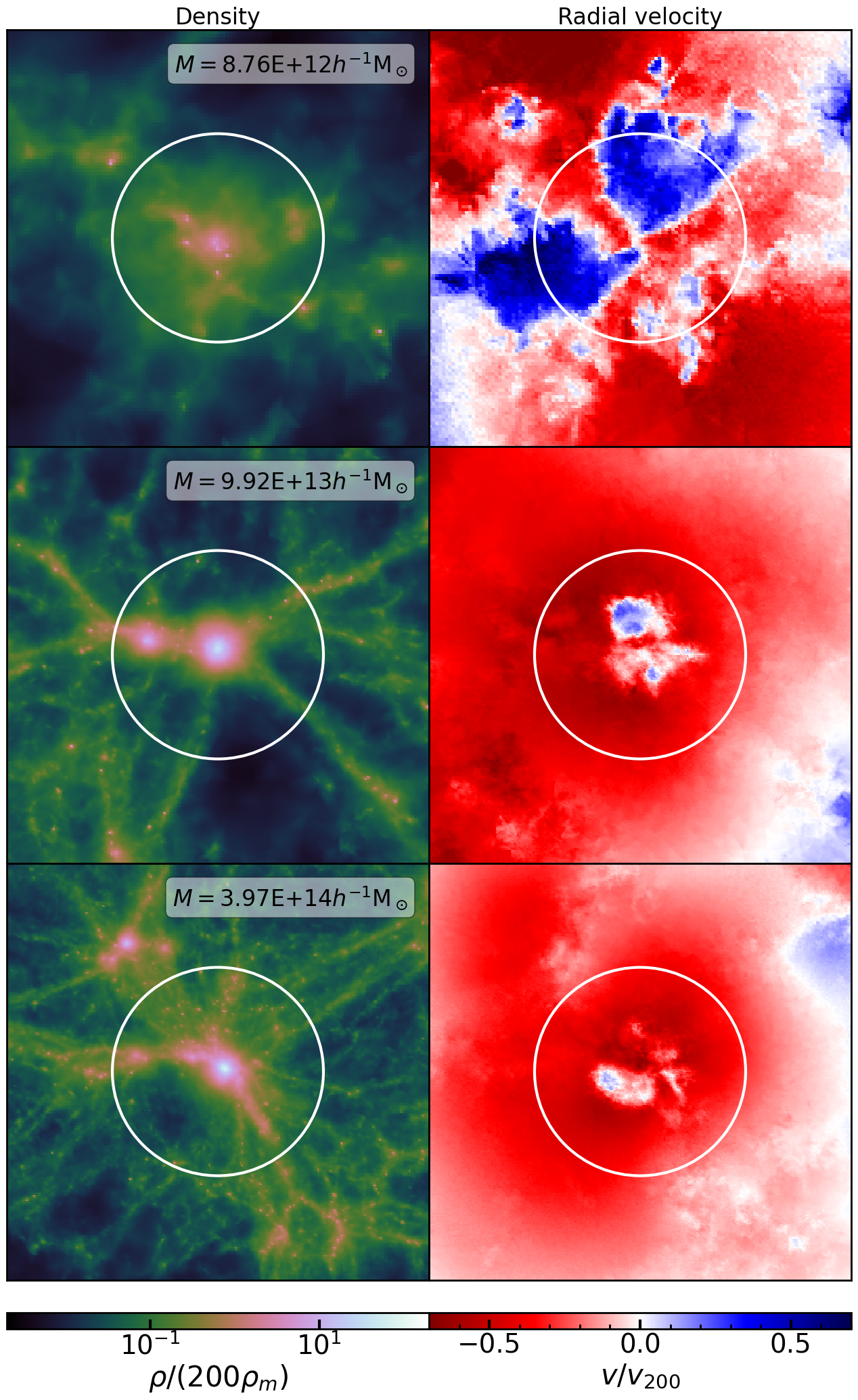}
        \label{fig:halo_example_gas}
    \end{subfigure}
    \caption{Density (left) and radial velocity (right) fields around three haloes of mass $M_{200}=8.7 \cdot 10^{12}, 9.92 \cdot 10^{13}, 3.97 \cdot 10^{14} h^{-1} \rm M_{\rm \odot}$ and radius $r_{200}= 0.50, 1.12, 1.77 h^{-1} \rm Mpc$, from top to bottom. Figure \ref{fig:halo_example_dm} and \ref{fig:halo_example_gas} display the respective dark matter and gas fields of the fiducial ($\rm fgas$) run. The panels show a region of $\pm 10 r_{200}$ around the halo centre, projecting a slice of thickness $4r_{200}$ in the $z$ direction. The white circles represent a $5 r_{200}$ radius. Positive (negative) radial velocities are shown in blue (red).}
    \label{fig:combined}
    
\end{figure*}

As an example, Figures \ref{fig:halo_example_dm} and \ref{fig:halo_example_gas} visualise three haloes of mass $M_{200} \approx 8.7 \cdot 10^{12} h^{-1} \rm M_{\odot}$ (upper row), $M_{200} \approx 9.92 \cdot 10^{13} h^{-1} \rm M_{\odot}$ (middle row) and $M_{200} \approx 3.97 \cdot 10^{14} h^{-1} \rm M_{\odot}$ (lower row) in the fiducial run. The left and right columns display density and radial velocity, respectively, where blue colours denote positive (outflowing) radial velocities, and red colours negative (inflowing) radial velocities, normalised to $\pm v_{\rm 200} = \pm \sqrt{GM_{\rm 200}/r_{200}}$. Figure \ref{fig:halo_example_dm} corresponds to the dark matter, and Figure \ref{fig:halo_example_gas} to the gas. We used the DTFE algorithm \citep{Cautun:2011} to estimate density and velocities on a grid and then projected a slice of $\pm 2 r_{200}$ along the line of sight direction. The cell size is $70, 80, 140 h^{-1} \rm kpc$ in each halo mass. A region of $\pm 10 r_{200}$ around the halo is shown in each case, and the white circle denotes the radius of $5 r_{200}$.

A quick visual inspection readily shows the impact of feedback on the density and velocity fields. Compared to dark matter, the most significant differences appear at the lowest mass haloes, where AGN feedback can overcome the halo's gravitational potential and efficiently redistribute gas well beyond the boundary of the halo. As a result, the gas is more extended than the dark matter. A complementary picture appears in velocity space: while the bulk of the dark matter is infalling (red) and gets virialised within the halo (white central region), AGN feedback events push gas with positive velocities out of the halo (blue bubbles). 
In the fiducial FLAMINGO model that is depicted in the figures, the outflows reach up to $\approx 6 r_{200}$ with velocities of the order of $v_{\rm 200}$ in the lowest mass halo. In the following section, we quantify such effects. 

\subsection{Density and radial velocity profiles}\label{sec:profiles}

\begin{figure*}
    \centering
    \includegraphics[width=0.8\linewidth]{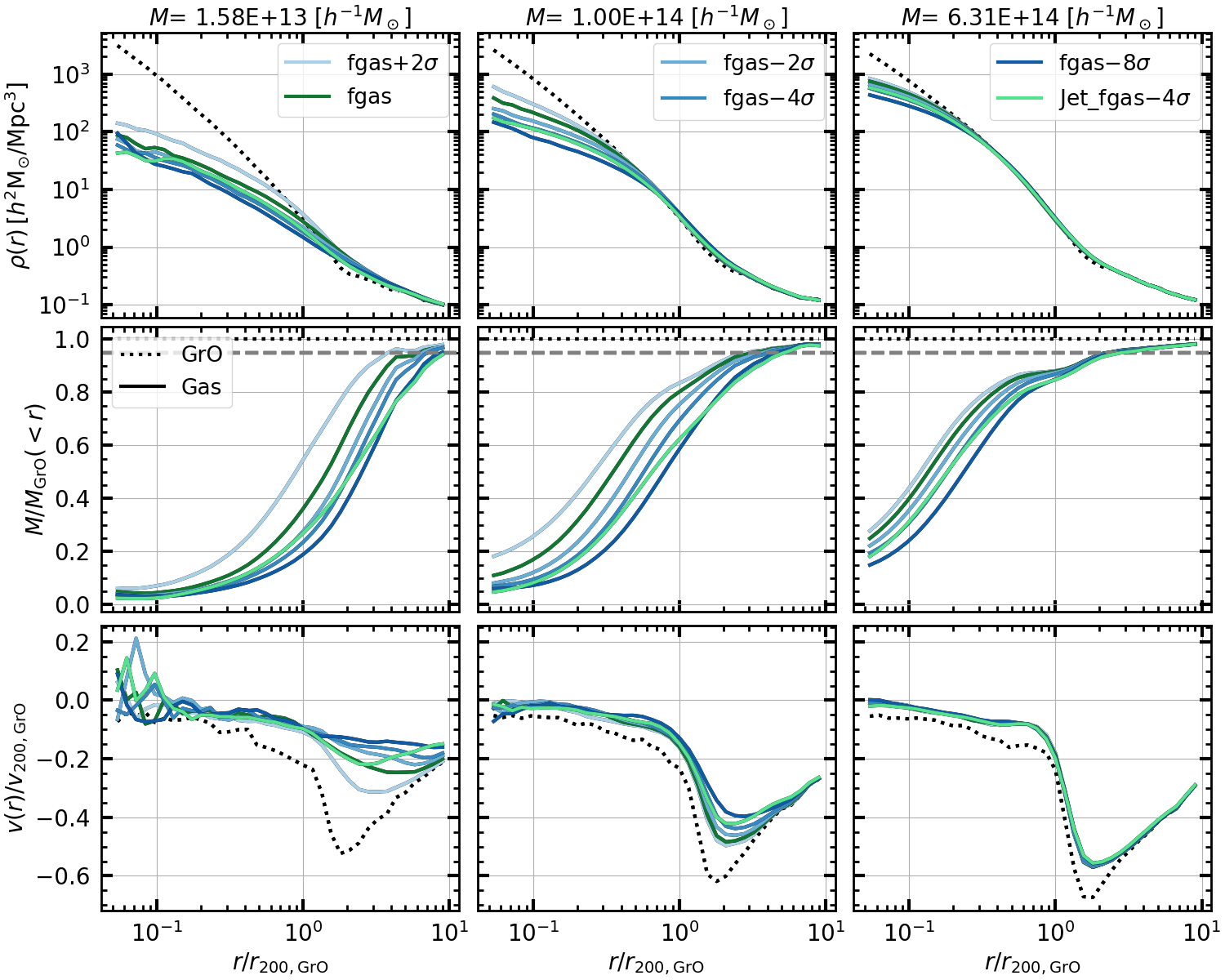}    
    \caption{Gravity-only (dotted black lines) and gas (solid lines) spherically averaged radial profiles around isolated haloes of three mass bins at $z=0$. \textit{Top row}: Density profiles, where the gas is normalised by the cosmic gas fraction. \textit{Middle row}: Enclosed mass profiles over the gravity-only enclosed mass, where the gas is normalised by the cosmic gas fraction. \textit{Lower row}: Radial velocity profiles over $v_{200}$.}
    \label{fig:profiles}
\end{figure*}

We compute spherically averaged density and radial velocity profiles of gas around haloes for different FLAMINGO models. The halo positions are defined as the minimum of the potential, while the halo velocity is the centre of mass velocity of the halo.  We crossmatch\footnote{For each halo in the gravity-only simulation, we find the closest halo in the full-physics run, imposing that their logarithmic mass difference cannot be larger than one and their distance cannot be larger than $3 r_{200}$. Then, we apply the same procedure starting from the haloes in the full-physics run. We consider only the pairs found in both directions to be cross-matched. After crossmatching haloes between the gravity-only and all FLAMINGO runs, we only select the gravity-only haloes that found a crossmatch in all FLAMINGO simulations.} the haloes in the gravity-only and full-physics runs so that we compare the same haloes among different models. Depending on the model, haloes may have lost some mass due to feedback, so we classify them in mass bins using their gravity-only mass. Moreover, we only consider isolated haloes. A halo is isolated if no halo more massive than itself is found within $20 r_{200}$. This is an important criterion mainly for the lowest mass bin, where many haloes fall towards a larger one. The merging signal dominates in those cases, and the density and velocity profiles change dramatically. As we are only interested in the smooth gas component rather than in the effects of feedback in the merging history of haloes, we impose a somewhat extreme isolation criterion \citep[for a similar approach, see][]{Lau:2015}. Finally, we select 100 crossmatched haloes in each mass bin and stack the profiles.

Figure \ref{fig:profiles} shows i) the density, ii) the ratio of the enclosed mass of gas and dark matter of the gravity-only run, and iii) the mean radial velocity profiles in three halo mass bins, increasing halo mass from left to right. The mean radial velocity is computed using all the particles that fall in a given spherical shell. The profiles are computed per halo, and then we take the average of the profiles of all haloes in each mass bin. Black dotted lines represent the gravity-only profile, and solid lines represent the gas. Gas density and mass profiles are normalised to their cosmic mass fraction $f_{\rm gas}$ for easier comparison. We checked that there are no noticeable differences between the dark matter on the full-physics run and that of the gravity-only run; hence, we only show the gravity-only result. Different colours denote different FLAMINGO models, as indicated by the legend. 

For all halo masses, stellar and AGN feedback push the gas away from the central region of the halo, lowering the gas density. In low mass haloes, the ejected gas reaches $r>r_{\rm 200,GrO}$. More than half of the gas is ejected and remains outside $r_{\rm 200,GrO}$, and only around $r \approx 10 r_{\rm 200,GrO}$ the enclosed gas-to-matter ratio approaches the cosmic value. At higher halo masses, feedback cannot overcome the gravitational potential of the halo, and the differences among models are restricted to the central regions, $r<r_{\rm 200, GrO}$. Yet, the mass distributions of gas and dark matter are not the same at the boundary of the halo. This is visible in the ratio of their enclosed masses at the highest mass bin: around $r \approx (0.6-1.0) \, r_{\rm 200, GrO}$ dark matter and gas seem to trace each other, but at $r>r_{\rm 200, GrO}$ there is more gas than dark matter, once normalised by the gas fraction.
\begin{figure}
    \centering
    \includegraphics[width=\columnwidth]{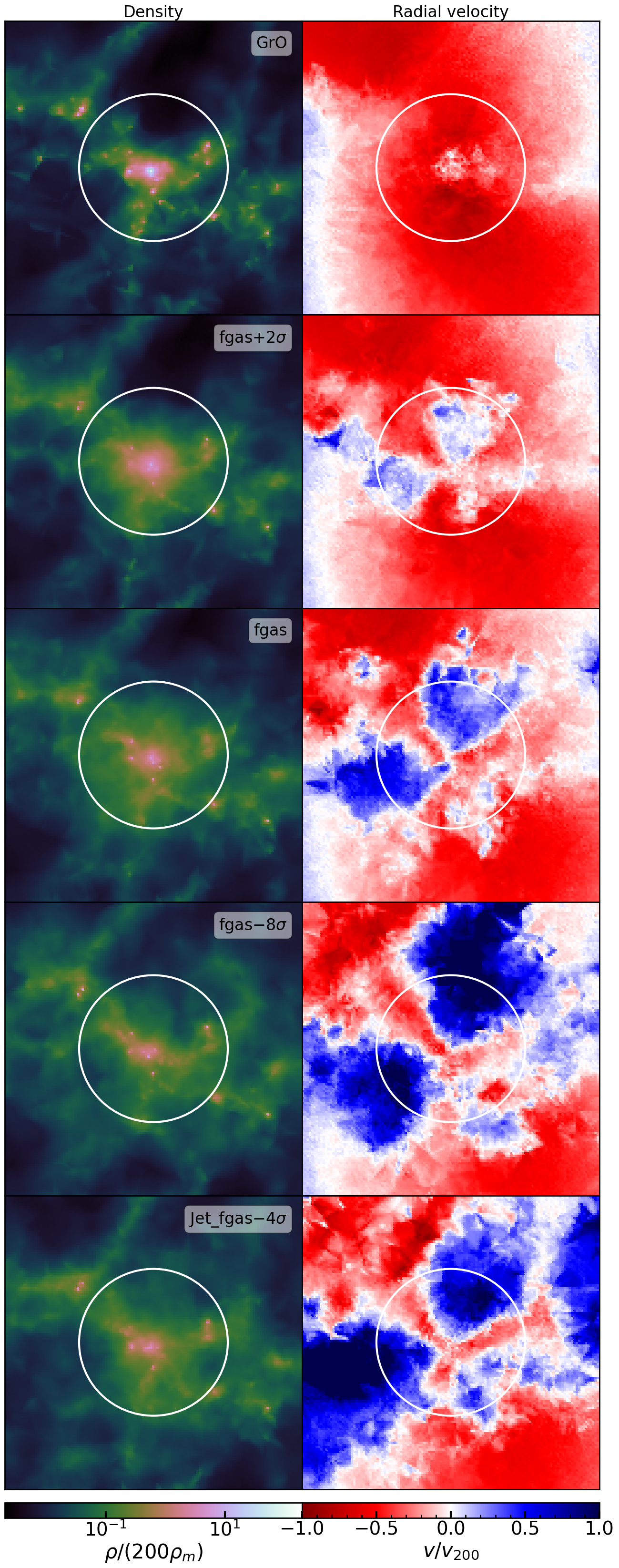}
    \caption{Density (left) and radial velocity (right) fields around the same halo of $M_{\rm GrO}= 1.00 \cdot 10^{13}h^{-1}\rm M_{\rm \odot}$ and $r_{200}=500 \, h^{-1}\rm kpc$. The upper panel shows the halo in the gravity-only simulation, while all the others show the gas component of the halo in different feedback scenarios, with increasing strength from top to bottom. The panels show a region of $\pm 10 r_{200}$ around the halo centre, projecting a slice of thickness $4r_{200}$. White circles represent a $5 r_{200}$ radius. Positive (negative) radial velocities are shown in blue (red).}
    \label{fig:halo_example_physics}
\end{figure}

Mean radial velocities show a consistent picture. In the highest mass bin, both gas and dark matter are being accreted by the halo with negative velocities. When they reach the boundary of the halo, the average velocity rapidly reaches zero. Dark matter and gas particles draw orbits around the halo centre, averaging inward and outward movements. On top of this, gas is collisional, and when it gets to the boundary of the halo, it feels ram pressure and is shocked. The combination of both effects slows the gas while the dark matter pierces through the halo without any resistance. As a consequence, gas velocities are less negative and converge faster to a mean of zero \citep[e.g.][]{Lau:2015}. Additionally, gas piles up at the boundary of the halo, creating the step-like feature in the enclosed mass profiles. Thus, in addition to using tSZ observations to measure the shock location in clusters \citep{Baxter:2021, Adhikari:2021, Anbajagane:2022,Towler:2024}, we speculate that kSZ observations could serve as a complementary tool for this purpose.

At lower halo masses, the average gas radial velocity profiles are shallower and model-dependent, becoming less negative the stronger the feedback (as parametrised through the gas fraction within $r_{500c}$). Relative to the underlying dark matter, on average, gas is accreted $50-80\%$ more slowly. The range of scales affected is wider too,  $\approx [2-10] r_{200,\rm GrO}$. It is not straightforward to understand this. There could be several things happening. A possibility is that, as feedback ejects gas outside the halo (more or less effectively, depending on the feedback strength), the infalling gas encounters different environments when it arrives at the boundary of the halo. Thus, in stronger feedback scenarios, the infalling gas finds ejected gas at larger radii and gets shocked, losing its kinetic energy, driving the average radial velocity towards zero faster -- i.e. by reducing the negative radial velocity (infalling) tail of the distribution. This is consistent with the discussion of \cite{Kwan:2024}, where they argued that the multistream region of the gas is more extended than dark matter's at low halo masses. Another possibility is that the positive velocities of outflows contribute in a non-negligible way out to large radii and drive the average velocity towards zero -- i.e. by increasing the positive radial velocity (outflowing) tail of the distribution. Most probably, it is a combination of both. In the next section, we explore the role of the latter.

Nevertheless, there are clear differences between the velocity profiles of \cite{Kwan:2024} and ours. First, the velocity profiles in \cite{Kwan:2024} have a minimum around the virial radius, and a maximum at slightly larger radii than the minimum. At larger radii, the velocity profiles drop again, reaching even more negative velocities than at the minimum, mainly in low-mass haloes. This feature is consistent with their low mass haloes not being isolated \textcolor{black}{- even if not shown, we have verified that we find the same trends if we drop the isolation criterium}. Close to the virial radius, the particles that are indeed falling to the halo dominate the profile. At larger radii, though, the low mass haloes themselves are falling into the potential of a more massive companion, which dominates over the small halo particles and creates misleading negative velocities "towards" the small halo in the centre (notice that the amplitude of the infall velocities scales roughly as $v_{200} \propto \sqrt{M_{200}/r_{200}}$). 

Second, \cite{Kwan:2024} find that the several BAHAMAS variations with different feedback strengths do not impact noticeably the velocity profiles. However, in FLAMINGO the changes between different feedback strengths can get to $40 \%$ differences around $2 r_{200}$ (lower left panel of Figure \ref{fig:profiles}). This difference is probably a consequence of \cite{Kwan:2024} not having isolated haloes: the particles of the larger halo dominate in the computation of the mean radial velocity, and the differences in the smooth component become invisible. \textcolor{black}{Finally, \cite{Kwan:2024} do not match haloes across different simulations when comparing their velocity profiles. This likely hides the impact of feedback on individual haloes too.}

\subsection{Outflows}

In Figure \ref{fig:halo_example_physics} we plot density and radial velocity fields around the same $M_{200}=1.00 \cdot 10^{13}h^{-1} \rm M_{\odot}$ halo: dark matter in the gravity-only run (upper row) and gas in several FLAMINGO models (from second to last row), with increasing feedback strength. The lower panels show the jet implementation of AGN feedback, while the other runs use the thermal implementation. A region of $\pm 10 r_{200}$ around the halo is shown, where $r_{200} = 500 h^{-1} \rm kpc$. Both densities and velocities share colour normalization in all cases. In particular, radial velocities are coloured in the range $\pm v_{\rm 200, GrO}$.

The outflows vary dramatically with feedback strength. On the one hand, the strength sets the scales the outflows reach: in the weakest feedback scenario of the FLAMINGO models (second row), they get out to $\approx 5 r_{200}$, while for the strongest feedback, they reach further than $\approx 10 r_{200}$. Notice that these outflows, identified by the radial velocity, extend to much larger scales than the high-entropy outflows previously found in simulations \citep{Tremmel:2019, Nobels:2022}, which are confined to $\approx 0.1 \rm r_{200}$. In SIMBA, the outflows reach $\approx 4 r_{200}$ \citep{Yang:2024}. On the other hand, the outflow speed also scales with feedback strength. In $\text{fgas}-8\sigma$, gas is outflowing with velocities comparable to $v_{200, \rm GrO}$, while in $\text{fgas}+2 \sigma$, gas is slower than $\approx 0.5 v_{200, \rm GrO}$. 

\subsubsection{Anisotropic feedback}

Feedback is not an isotropic process. The radial velocities clearly show a biconical bubble structure of the outflows. Despite the energy injection in the thermal AGN implementation being isotropic \citep{Booth:2009}, the gas takes the path of least resistance and expands towards low-density regions, creating buoyant bubbles \citep[e.g.][]{Tremmel:2019, Nobels:2022}. In the jet implementation, the injection is done by kicking two particles in the direction of two cones
with 7.5° opening angles around the BH spin axis \citep{Husko:2022, Schaye:2023}, but that does not prevent them from losing the dependence of the initial direction at large scales and forming buoyant bubbles \citep{Husko:2024}. In fact, visually, the bubbles in jet and thermal implementations do not show obvious differences. 

The outflows transport gas from the halo centre towards the outskirts, making the gas less dense in the centre for increasing feedback strength. Presumably, due to the anisotropic structure of the outflows, the redistribution of gas is also anisotropic, transporting more gas to low-density regions than to high-density ones and making the overall gas distribution more spherical. In Figure \ref{fig:halo_example_physics}, the visual comparison of gas density around $\approx 5 r_{200}$ with increasing feedback strengths suggests so. We speculate that stacking kSZ measurements based on orientation could provide a new method to constrain feedback effects. 

Similar structures have been found in other hydrodynamic simulations with different subgrid models. In IllustrisTNG and EAGLE \citep{Peroux:2020, Truong:2021a, Truong:2021, Truong:2023} the outflows were typically confined to the halo, while in SIMBA they reach $\approx 4 r_{200}$ with the jet-variant \citep{Yang:2024}. These works define the major and minor axis of the galaxy, and analyse the anisotropy of feedback according to those directions. They analyse several gas properties, such as metallicity, temperature, X-ray luminosity, velocity and densities along the minor and major axis directions. All properties except the density are enhanced in the minor axis direction compared to the major axis due to feedback activity. Hence, they propose oriented X-ray observations to detect such bubbles in the real universe, and forecast that they should lie on the detection threshold of eROSITA \citep{Truong:2021,Truong:2021a,Yang:2024}. Alternatively, \cite{Martin-Navarro:2021} reported a complementary signature in anisotropic quenching of galaxies in Sloan Digital Sky Survey (SDSS) data \citep{Ahn:2014}. \textcolor{black}{However, later works showed that the anisotropic quenching is a prediction of hierarchical structure formation rather than being an indicator of anisotropic AGN feedback \citep{Karp:2023}}. 

Whereas previous works find a decrease in gas density in the outflow direction, we argue that the visual inspection of Figure \ref{fig:halo_example_physics} suggests the opposite. There could be several possibilities for this discrepancy: on the one hand, we define the anisotropy with respect to the large-scale density distribution rather than the galaxy's orientation. However, those two are statistically correlated in observations \citep{Rodriguez:2022} and simulations \citep{Rodriguez:2023}. On the other hand, the scales that we are probing are also different: typically within the halo vs at several virial radii. We will further explore and quantify this in future works. 

\subsubsection{Mass and radial scaling of outflows}

\begin{figure}
    \centering
    \includegraphics[width=\columnwidth]{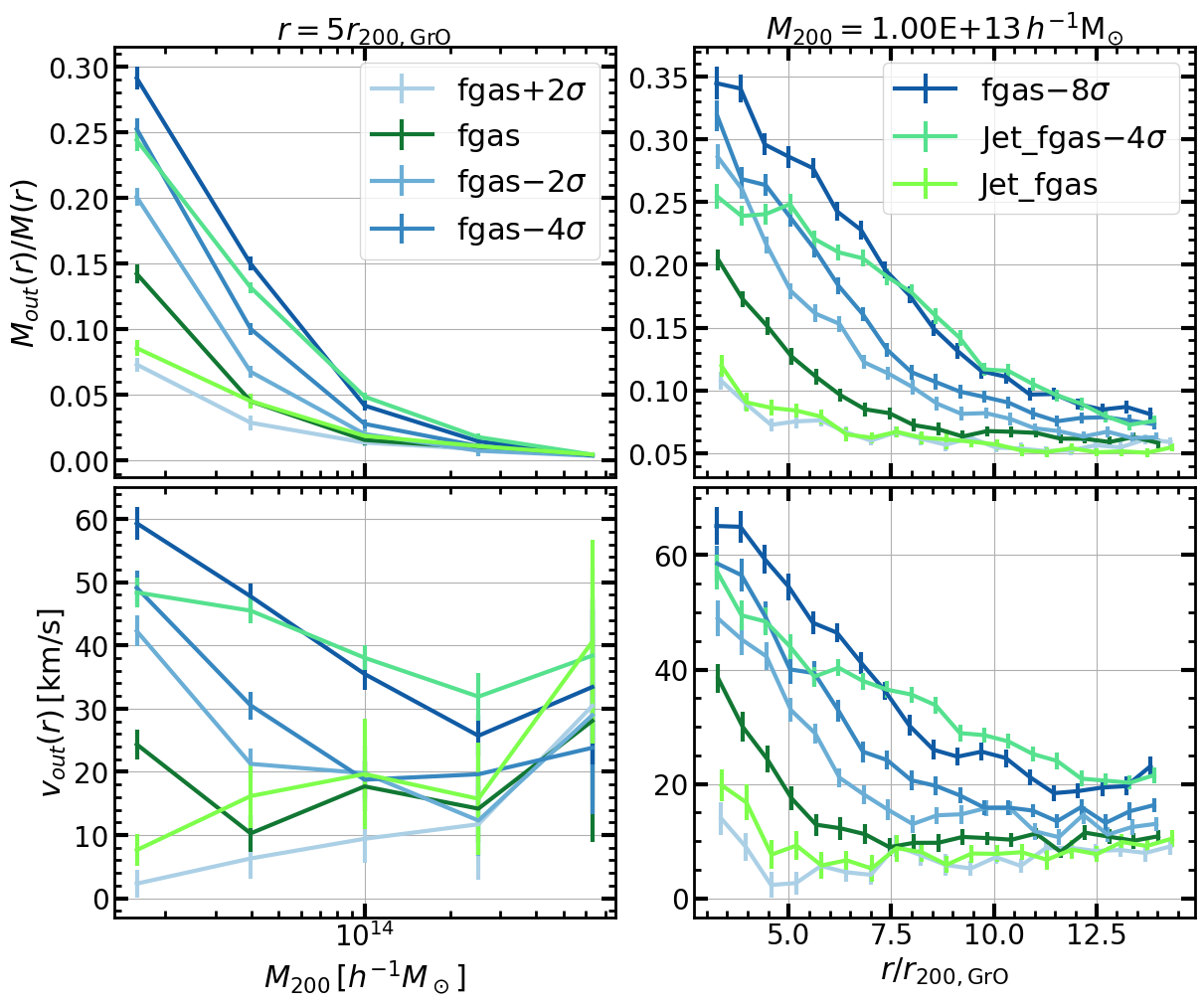}
    \caption{Radial and halo mass scaling of gas outflows at $z=0$. \textit{Upper row}: Gas mass fraction in outflows, defined as the ratio of the outflowing gas mass over total gas mass in a given radial shell. \textit{Lower row}: Mean velocity of outflows in a given radial shell. \textit{Left column}: Halo mass scaling of outflows at fixed radius $r=5r_{\rm 200, GrO}$. \textit{Right column}: Radial scaling of outflows at fixed halo mass $M_{200}=10^{13}h^{-1} \rm M_{\odot}$. The averages are taken over 100 isolated haloes in each mass bin between $M_{200} \in (10^{13},10^{15}) h^{-1} \rm M_{\odot}$, and the error bars denote the standard deviations. The colours stand for different FLAMINGO models.}
    \label{fig:outflows}
\end{figure}

To quantify the relative importance of outflows, we need criteria that identify them. One could define any particle with positive radial velocities as outflowing, as done in \cite{Ayromlou:2023}. However, this simple criterion makes the interpretation of the results more complex. Because both dark matter and gas are virialised within the halo, and they have some velocity dispersion associated with them, this criterion identifies as outflows $\approx 50 \%$ of the material within the halo, both in gas and dark matter components (white region - average velocity 0 - in the centre of the upper right panel in Figure \ref{fig:halo_example_physics}). Similarly, the material far away from the halo, probably falling into another halo (boundaries of the upper right panel in Figure \ref{fig:halo_example_physics}) is also considered as outflow\footnote{Notice that because we are using comoving coordinates with peculiar velocities, the Hubble flow is not included.}. Thus, at least in the small and large-scale limits, the results would not have a clear meaning.

Our approach is to define outflowing or inflowing regions and then tag the particles in those regions accordingly. We do so by using only the kinematic information of gas and dark matter particles. Using the DTFE algorithm, we compute the radial velocity in a grid around each halo, with a cell size of $l = 30 r_{200} / 128$. From the lowest (highest) mass haloes in our sample, this gives a cell size of $120 h^{-1}\rm kpc$ ($420 h^{-1}\rm kpc$). Unlike other methods, the DTFE algorithm provides a value for the velocity at each point in space. Additionally, computing velocities on a grid by construction removes a part of the intrinsic velocity dispersion. Finally, a grid cell is considered as outflowing if the following conditions are satisfied: 
\begin{equation}\label{eq:out_sel_0}
\begin{aligned}
    V_{\rm gas}(x,y,z) >  0 , \\
    V_{\rm dm}(x, y,z) < 0,
\end{aligned}
\end{equation}
where $V_{\rm gas}(x,y,z)$ and $V_{\rm dm}(x,y,z)$ are the radial velocity grids of gas and dark matter in the full-physics run. Once we have identified the outflowing regions, we tag the particles that fall in those regions as outflowing. Finally, we describe outflows with the mass that they carry ($M_{\rm out}$) and their median velocity ($v_{\rm out}$) in a given radial bin. 

The results are shown in Figure \ref{fig:outflows}. We take the mean and standard deviation of 100 isolated haloes in each mass bin. We display the amount of mass in outflows over the total gas mass (upper row) and its velocity (lower row) in a given radial shell. In the left column, we compute this quantity as a function of halo mass at a fixed radius, while in the right column, we keep the halo mass fixed and change the radius. The different colours denote different FLAMINGO models. 

Notice that even though the trends are robust to the details of the selection criteria, the exact numbers do change. For instance, by setting the minimum velocity to be considered as outflowing to 0 (Eq. \ref{eq:out_sel_0}), we effectively get an upper limit on the mass in outflows and a lower limit on their velocities. \textcolor{black}{Furthermore, feedback from infalling haloes is also expected to contribute to the total outflowing gas. The isolation criterion used in the halo selection likely reduces this contribution. As a result, we may underestimate the relative significance of outflows for haloes of a given mass.} Thus, we mainly consider the overall trends and relative differences between FLAMINGO models. Additionally, as mentioned above, in virialised regions, it is hard to reliably distinguish outflowing gas from regions with positive radial velocities within their orbit using only kinematic information. Therefore, we make a conservative cut and show the results only at several virial radii.

\begin{figure}
    \centering
    \includegraphics[width=0.9\linewidth]{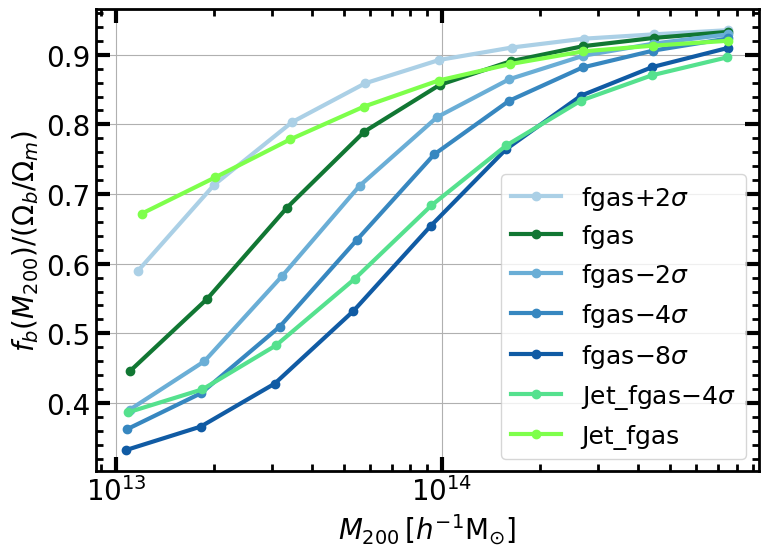}
    \caption{The average halo baryon fraction at $r_{200}$ as a function of halo mass at $z=0$. The results are shown for isolated haloes with $M_{200}>10^{13}h^{-1}\rm M_{\odot}$. The colours denote different FLAMINGO models.}
    \label{fig:Mh_frac}
\end{figure}

Overall, the selection shows the expected behaviour with halo mass and distance from the halo. At fixed feedback strength, outflows carry relatively more mass at low halo masses (top left) and closer to the halo (top right). As we go further from the halo, the outflows slow down (bottom right). It is not clear how the velocities scale with halo mass (bottom left): in some runs, they become slower in larger haloes ($\text{fgas}-8\sigma$, $\text{Jet\_fgas}-4\sigma$), in others they get faster ($\text{fgas}-2\sigma$,$\text{Jet\_fgas}$), and in others, they are nearly constant. This is highly dependent on the distance from the halo centre too. 

The stronger the feedback, the more gas is outflowing and the higher its velocity. This generally agrees with the qualitative trends inferred from Figure \ref{fig:halo_example_physics}. For low halo masses, where feedback is most effective, the percentage of outflowing gas over total gas mass can vary by a factor of a few between the weakest ($\text{fgas}+2\sigma$) and strongest ($\text{fgas}-8\sigma$) feedback runs at a fixed radius. Moreover, outflowing gas reaches larger scales in stronger feedback scenarios. 



Interestingly, the radial and halo mass profile of the outflows in the $\rm Jet$ runs seems to differ from those in thermal AGN implementations. In particular, compared to the thermal AGN runs, the outflows in $\text{Jet\_fgas}-4\sigma$ are relatively less massive and slower in the inner regions (low halo masses), becoming relatively more massive and faster in the outer regions (high halo masses). This suggests that the radial profile and halo-mass scaling of outflowing gas (and therefore, the impact of feedback on the total gas distribution around haloes) depend noticeably on the AGN feedback implementation scheme\textcolor{black}{, even for models calibrated to the same cluster gas fractions and galaxy mass function.} 

In Figure \ref{fig:Mh_frac}, we show the average halo baryon fraction within $r_{200}$ as a function of halo mass for different FLAMINGO models. We compute the halo baryon fraction of isolated haloes with $M_{200} \in (10^{13}-10^{15}) h^{-1} \rm M_{\odot}$ in 10 logarithmically spaced mass bins, randomly selecting 1000 haloes in each bin. We select only crossmatched haloes, and classify them in mass bins using their gravity-only mass, as described in Section \ref{sec:profiles}. We compute the halo baryon fraction as the baryonic to total halo mass within $r_{200}$. Notice that this plot is similar to Fig.7 of \cite{Kugel:2023}; that is, it features the gas fractions as a function of the halo mass of the data used to calibrate the simulations. \textcolor{black}{Due to the limited box size of the calibration runs, only halo masses up to $M_{500} = 2.5 \cdot 10^{14} \, \rm M_{\odot}$ were used in the calibration.} Hence, the halo mass scaling is at least partially driven by the calibration data. 

For the same calibration data, it becomes evident that Jets are comparatively less efficient at lowering the baryon fraction within $r_{200}$ at lower halo masses, in agreement with Figure \ref{fig:outflows}. However, as visible in the enclosed mass profiles of Figure \ref{fig:profiles}, the gas ejected by $\text{Jet\_fgas}-4\sigma$ reaches scales as large as the ejected gas in $\text{fgas}-8\sigma$. This is a simulation prediction: at a given halo baryon fraction, the extent of feedback differs for the thermal and jet AGN feedback implementations.

Complementary information is provided by previous works analysing the reach of the ejected gas in several hydrodynamical simulations \citep{Tollet:2019, Davies:2019, Angelinelli:2022, Sorini:2022, Ayromlou:2023b}. In particular, \cite{Ayromlou:2023b} found that different hydrodynamical simulations redistribute gas to different scales at fixed halo baryon fraction within $r_{200}$ -- the so-called closure radius. A comprehensive analysis of the closure radius in FLAMINGO simulations will be presented in Denison et al. (in prep).

\section{Summary and conclusions}\label{sec:conclusions}

In this paper, we have employed the state-of-the-art FLAMINGO simulations \citep{Schaye:2023} to study the impact of hydrodynamical forces and galaxy formation physics on the density and velocity fields of cosmic gas. This suite has been calibrated with observed gas and stellar fractions within galaxy clusters and groups \citep[][]{Kugel:2023}, and is therefore particularly suited for this study. 


In the first part of the paper, we have explored several clustering statistics of cosmic gas – such as the density power spectra, velocity-divergence power spectra and mean pairwise velocities – and characterised the differences relative to expectations from gravity-only simulations. Afterwards, we have focused on the properties of the gas around haloes of mass between $M_{200} \in (10^{13},10^{15}) h^{-1} \rm M_{\odot}$, disentangling different physical processes at play and quantifying the effect of subgrid parameters and AGN feedback implementations.

Our main findings are the following:
\begin{itemize}
    \item On large scales, the gas velocity is identical to that of dark matter (Fig.~\ref{fig:power-spectra-vdiv}). However, the gas density field is biased low relative to that of dark matter even on scales as large as $k=0.01h / \rm Mpc$ (Fig.~\ref{fig:power-spectra}). At $z=0$, this bias is $b^2 \approx 5\%$ in the power spectrum, with a maximum amplitude of $b^2 \approx 8\%$ at $z \approx 1$. Indirectly, this also creates a large-scale bias in the mean pairwise velocities of gas, which could be misinterpreted as a sign of gas moving more slowly than dark matter at scales as large as $r \approx 10 h^{-1}\rm Mpc$ (Fig. \ref{fig:pairwise-velocities}). We show that the bias is almost independent of the gas fraction in clusters, but does noticeably depend on the stellar mass function (Fig. \ref{fig:power-spectra}). We interpret this as a consequence of star formation in dense and clustered regions, leaving gas preferentially in less dense regions.
    \item On small scales, gas is greatly affected by non-gravitational forces. The power spectrum of gas is strongly suppressed relative to the gravity-only case (Fig. \ref{fig:power-spectra}), as are the potential flows (Figs. \ref{fig:power-spectra-vdiv} and \ref{fig:vdiv_dens}) and the mean pairwise velocities (Fig. \ref{fig:pairwise-velocities} and \ref{fig:pairwise-velocities-cross}). While this is arguably the combined effect of several non-gravitational processes, it strongly correlates with the feedback strength of the FLAMINGO simulations.
 

    \item In agreement with previous work, we find that in the FLAMINGO simulations, AGN feedback reshapes the density field by pushing gas away from haloes, being most effective in group-scale haloes with $M_{200} \approx 10^{13}h^{-1} \rm M_{\odot}$ (Fig. \ref{fig:profiles}). 

    \item  Gas radial velocity profiles differ from the gravity-only profiles in two ways. First, in massive galaxy clusters, gas gets shocked at the halo boundary and loses kinetic energy, driving its mean radial velocity towards zero faster than dark matter's. Second, for group-scale haloes ($M_{200} \approx 10^{13}h^{-1} \rm M_{\odot}$), the radial velocity profiles noticeably depend on feedback at radial distances as large as $\approx 10\,r_{200, \rm  GrO} \approx 5 h^{-1} \rm Mpc$ (lower panels of Fig. \ref{fig:profiles}). 

    \item Feedback-dependent changes in the density and radial velocity profiles can be explained mainly in terms of large outflows sourced by the accretion onto supermassive black holes (Figs. \ref{fig:combined} and \ref{fig:halo_example_physics}). Irrespective of thermal or jet implementations of the AGN feedback, the surrounding gas eventually expands, creating buoyant bubbles that can reach tens of virial radii with radial velocities comparable to the escape velocity of the halo (for a visual impression, c.f. Fig. \ref{fig:halo_example_physics}). These bubbles show a clear biconical structure, implying that in FLAMINGO \textcolor{black}{AGN feedback becomes anisotropic on resolved scales.}      
    
    \item The fractional mass in outflows is larger at lower halo masses and near the halo boundary, whereas it becomes less important at larger distances (Fig. \ref{fig:outflows}). Baryonic effects on density and velocity scale with the AGN feedback strength. 

    \item The radial profile and halo mass scaling of outflows is sensitive to the AGN feedback implementation (Fig. \ref{fig:outflows}). Jets become less efficient than thermal AGN at lowering the halo gas fractions within $r_{200}$ for lower halo masses (Figs. \ref{fig:profiles}, \ref{fig:outflows} $\&$ \ref{fig:Mh_frac}). However, they redistribute gas to larger distances (Figs. \ref{fig:profiles},\ref{fig:outflows} $\&$ \ref{fig:Mh_frac}).
\end{itemize}

In summary, the connection between cosmic gas and dark matter is modulated by galaxy formation processes across a wide range of scales. In particular, gas around group-size haloes is especially susceptible to feedback processes, significantly impacting the density and velocity fields out to large scales. Thus, in principle, measurements of the gas density and velocity in the outskirts of these haloes are reliable proxies for AGN feedback. 

To this end, the kSZ effect is a powerful probe of the gas distribution around group-size haloes. The kSZ signal is proportional to the line-of-sight integral of the gas momentum, so it directly measures the gas density and velocity fields without the need to model the gas pressure or temperature (unlike thermal Sunyaev-Zeldovich or X-ray measurements). Thus, kSZ analyses will provide independent constraints on AGN feedback, which can be compared and combined with other complementary datasets, informing hydrodynamic simulations. 
This will be highly valuable for weak lensing surveys, as it helps isolate the effects of galaxy formation on the matter field and allows for the extraction of robust cosmological information.


\begin{acknowledgements}
      We thank Daniele Sorini for helpful comments. LO acknowledges the support of ”la Caixa” Foundation (ID 100010434) for the fellowship with code LCF/BQ/DR21/11880028. REA acknowledges support from project PID2021-128338NB-I00 from the Spanish Ministry of Science and support from the European Research Executive Agency HORIZON-MSCA-2021-SE-01 Research and Innovation programme under the Marie Skłodowska-Curie grant agreement number 101086388 (LACEGAL). \textcolor{black}{IGM acknowledges the support of the Science and Technology Facilities Council (grant number ST/Y002733/1).
      This project has received funding from the European Research Council (ERC) under the European Union’s Horizon 2020 research and innovation programme (grant agreement No 769130).}
\end{acknowledgements}

%
%
\bibliographystyle{aa} 
\bibliography{bibliography} 

\begin{appendix}
\section{Redshift evolution of gas bias}\label{sec:gas_bias_z}

\begin{figure}
    \centering
    \includegraphics[width=\linewidth]{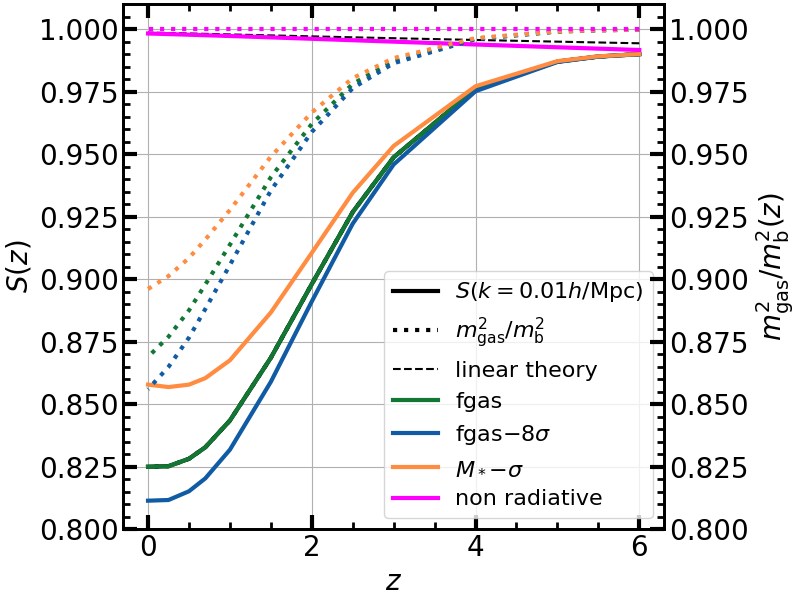}
    \caption{Split of the gas bias on its components: suppression of the gas power spectra ($S(k=0.01 h /\rm Mpc)$) and the evolution of the total gas to baryon mass squared ($m_{\rm gas}^2/ m_{\rm b}^2$). If the gas was not biased with respect to the gravity-only field, the solid lines should lie on top of the dotted lines. \textit{Solid lines}: Redshift evolution of the suppression of gas power spectrum, as defined by the ratio of gas and gravity-only power spectra at $k=0.01 h \rm Mpc$. The gas power spectrum is normalised to the total baryon mass. In dashed lines, we plot the linear theory prediction for the baryon suppression. \textit{Dotted lines}: Redshift evolution of the gas to baryon mass fraction squared, defined as the total mass in gas over total baryon mass. The colours denote different FLAMINGO runs.}
    \label{fig:ps_redshift_appendix}
\end{figure}

In this appendix, we further explore the redshift evolution of the gas bias. In Figure \ref{fig:ps_redshift_appendix} we represent the gas bias, splitted on two components. In solid lines, we show the suppression of gas power spectra relative to gravity-only power spectra evaluated at $k=0.01 h / \rm Mpc$. Unlike in Section \ref{sec:dens_ps_large}, we normalise gas power spectra relative to the baryon mass. In dotted lines, the total gas mass is relative to the total baryon mass squared. That is, the gas bias, as shown in the right panel of Figure \ref{fig:power-spectra}, is given by the ratio of the two (solid/dotted) -- if there was no bias, the dotted and solid lines should overlap. The colours represent different FLAMINGO runs, as indicated by the legend. 

First, let's focus on the non-radiative run. Its gas mass fraction is always one because it does not form stars. However, at high redshift gas shows a small bias, consistent with the linear theory prediction coming from the different primordial clustering of baryons and dark matter (black dashed line) \citep[for a detailed discussion, see][]{Angulo:2013}.

The rest of the FLAMINGO runs do form stars. Thus, the fraction of their gas-to-baryon mass ratio decreases (dotted lines), starting around $z \approx 4-5$. Approximately, that is the time when the suppression in the gas power spectrum starts deviating from the linear theory expectation for the baryon fluid (coloured solid lines vs black dashed line). The ratio of both gives the gas bias (right panel of Figure \ref{fig:power-spectra}), increasing as more stars form in biased regions of the gas field at lower redshifts.

At low redshift $z \approx 1$, gas bias turns and becomes \textit{less} biased (right panel of Figure \ref{fig:power-spectra}). In Figure \ref{fig:ps_redshift_appendix}, we see the reason for that: at lower redshifts than $z \approx 1$, stars still form, and the gas-to-baryon mass ratio keeps decreasing. However, the ratio of the gas and gravity-only power spectra is flattened.

We interpret this again as the consequence of the evolution of stellar formation. At lower redshift, the new stars form in less biased regions compared to high redshifts, just due to the redshift evolution of structure formation -- at lower redshift, more and more "common" (less biased) peaks cross the density barrier for collapse or star formation. This was already theorised in early works of galaxy bias \citep{Fry:1996, Tegmark:1998} and found in simulations in \cite{Springel:2018}. The indirect consequence is that the gas field also becomes less biased.

\section{Mean pairwise velocity in linear theory}\label{sec:pairwise_derivation}

In this appendix, we will derive the linear theory expression at leading order of the mean pairwise velocity, defined by the following equation
\begin{equation}
    \langle  \omega (r) \rangle  = \frac{\langle (1+\delta_1)(1+\delta_2)(\vec{v}_2 - \vec{v}_1)\rangle  \cdot \hat{r}}{\langle (1+\delta_1)(1+\delta_2)\rangle },
\end{equation}
where $\vec{r} = \vec{r}_2-\vec{r}_1$ and the average is over all particle pairs at separation $|\vec{r}|=r$. Doing the multiplication,
\begin{equation}
     \langle  \omega (r) \rangle  = \frac{\langle \vec{v}_2-\vec{v}_1+\delta_2\vec{v}_2-\delta_1\vec{v}_1+\delta_1\vec{v}_2-\delta_2\vec{v}_1+\delta_1\delta_2(\vec{v}_2-\vec{v}_1)\rangle \cdot \hat{r}}{\langle 1+\delta_1+\delta_2+\delta_1\delta_2\rangle }.
\end{equation}
Let's study each term:
\begin{itemize}
    \item $\langle \vec{v}_1\rangle =0$ and $\langle \vec{v}_2\rangle =0$ because otherwise there would be a net motion.  
    \item $\langle \delta\rangle =\langle (\rho -\langle \rho\rangle )/\langle \rho\rangle \rangle =0$.
    \item $\delta_1\delta_2(\vec{v}_2-\vec{v}_1)$ is neglected at $O(3)$. 
    \item $\delta_1\vec{v}_1$ and $\delta_2\vec{v}_2$ have no real part in linear theory. This is clear when we change $\vec{r}_2$ by $\vec{r}_1$ in Equation \ref{eq:eq_ref0}.
    \item $\frac{1}{\langle 1+\delta_1\delta_2\rangle } \approx \langle 1 - \delta_1\delta_2\rangle  \approx 1$ at $O(3)$.
\end{itemize}
Thus, at leading order $O(2)$, the mean pairwise velocity reduces to
\begin{equation}
    \langle  \omega(r) \rangle  = [\langle \delta_1\vec{v}_2\rangle  - \langle \delta_2\vec{v}_1\rangle ]\cdot \hat{r}.
\end{equation}
Let's expand the first term in the equation. Writing the density and velocity in Fourier space,
\begin{equation}
    \langle \delta_1\vec{v}_2\rangle  =\biggl<  \int{d^3\vec{k} \tilde{\delta}_{k}\exp{(i\vec{k}\cdot \vec{r}_1)}  } \left[ \int{d^3\vec{k}' \tilde{\vec{v}}_{k'}\exp{(i\vec{k}'\cdot \vec{r}_2)} \delta_D(\vec{k}-\vec{k}') } \right] ^*\biggr>
\end{equation}
Applying the Dirac delta:
\begin{equation}
     \langle \delta_1 \vec{v}_2\rangle  = \biggl< \int{d^3\vec{k} \tilde{\vec{v}}_k^* \tilde{\delta}_{k} \exp{(-i\vec{k}\cdot \vec{r})}}\biggr>
\end{equation}
Moreover, in linear theory
\begin{equation}\label{eq:vel_dens_lin}
    -i \vec{k} \cdot \tilde{\vec{v}} = \tilde{\delta} a f H,
\end{equation}
where $a$ is the expansion factor of the Universe, $f$ is the linear growth rate, and $H$ is the Hubble parameter. We replace $\tilde{\delta}$,
\begin{equation}\label{eq:eq_ref0}
      \langle \delta_1 \vec{v}_2\rangle  = \frac{1}{afH} \biggl< \int{d^3\vec{k} (-i\vec{k}\cdot\tilde{\vec{v}}_k) \tilde{\vec{v}}_k^* \exp{(-i\vec{k}\cdot \vec{r})}}\biggr>
\end{equation}
As only the part in the radial direction survives, we can write $\tilde{\vec{v}}_k = \tilde{v}_k\hat{r}$. Thus, $\vec{k} \cdot \tilde{\vec{v}}_k^* = k \tilde{v}_k^* \cos{\theta}$, $\theta$ being the angle between $\hat{r}$ and $\hat{k}$. Besides, we can decompose differential $d^3\vec{k}$ 
\begin{equation}
    \int {d^3\vec{k}} = \int_0^{2\pi}{d\phi} \int_0^{\pi}{d\theta \sin{\theta}} \int_0^{\infty}{ dk k^2} = -2\pi \int_{-1}^{1}{d\cos{\theta}}\int_0^{\infty}{dk k^2},
\end{equation}
Putting everything together, we get
\begin{equation}
    \langle \delta_1 \vec{v}_2\rangle  = \frac{2 \pi \hat{r}}{afH}\int_0^{\infty} {dk k^2 \langle    \tilde{v}_k \tilde{v}_k^*\rangle } \int_{-1}^1{d\cos{\theta} [ik\cos{\theta}] \exp{(-ikr\cos{\theta})}} 
\end{equation}
which can be recasted into the matter power spectrum in linear theory (see Equation \ref{eq:Pvel_Pdens}),
\begin{equation}
   \langle \delta_1 \vec{v}_2\rangle  =  2 \pi afH \hat{r} \int_0^{\infty} {dk P(k)} \int_{-1}^1{d\cos{\theta} [ik\cos{\theta}] \exp{(-ikr\cos{\theta})}} 
\end{equation}
Expanding the exponential
\begin{equation}
    \exp{(-i\vec{k}\cdot{\vec{r}})} = \exp{(-ikr \cos{\theta})} = \cos(kr\cos{\theta})-i\sin(kr\cos{\theta}),
\end{equation}
and saving only the real part, we get
\begin{equation}
    \langle \delta_1 \vec{v}_2\rangle  = 2\pi afH \hat{r}\int_0^{\infty} {dk kP(k)}  \int_{-1}^1{d\cos{\theta} \cos{\theta} \sin(kr\cos{\theta})}
\end{equation}
Finally, we do a change of variables $u = \cos \theta$:
\begin{equation}
        \langle \delta_1 \vec{v}_2\rangle =  2\pi afH \hat{r} \int {dk kP(k)}  \int_{-1}^1{du u \sin(kru)}
\end{equation}
Doing the integral on $u$,
\begin{equation}
   \langle \delta_1 \vec{v}_2\rangle  =  -4 \pi afH \hat{r}\int_0^{\infty} {dk kP(k) \left[ \frac{\sin{kr}}{(kr)^2}-\frac{\cos(kr)}{kr}\right]}
\end{equation}
where, introducing the Bessel function definition
\begin{equation}
   \langle \delta_1 \vec{v}_2\rangle  = -4 \pi afH \hat{r}\int_0^{\infty} {dk kP(k) j_1(kr)}
\end{equation}
Notice that, $\langle \delta_1 \vec{v}_2\rangle = - \langle \delta_2 \vec{v}_1\rangle $ due to the imaginary number in equation \ref{eq:vel_dens_lin}. Thus,
\begin{equation}
    w(r)\hat{r} =\langle \delta_1 \vec{v}_2\rangle -\langle \delta_2 \vec{v}_1\rangle = 2\langle \delta_1 \vec{v}_2\rangle 
\end{equation}
So,
\begin{equation}
    w(r) = -8 \pi afH \int_0^{\infty} {dk kP(k) j_1(kr)}
\end{equation}
\end{appendix}
\end{document}